\documentclass[10pt, journal, twocolumn]{IEEEtran}

\usepackage{color,graphicx,amsmath,amssymb,amsthm,epsfig,mathrsfs,cite,bm,graphics}
\usepackage{caption}
\usepackage{algpseudocode}
\usepackage{algorithmicx}
\usepackage{algorithm} 
\usepackage[utf8]{inputenc}
\usepackage{amsmath}
\usepackage{amsfonts}
\usepackage{amssymb}
\usepackage{placeins}
\usepackage{graphicx}
\usepackage{bbm}
\usepackage{graphicx,graphics,subcaption}
\usepackage{url}

\usepackage{tikz}
\usetikzlibrary{automata, positioning, arrows,shapes.multipart}
\usepackage{amsmath,amsthm,bm}
\usepackage{amssymb}
\newcommand{\norm}[1]{\left\lVert#1\right\rVert}
\usepackage{xcolor}
\usepackage{graphics,graphicx,color,epsfig}
\usepackage{tikz}
\usetikzlibrary{automata, positioning, arrows,shapes.multipart,fit}
\usepackage{pgfplots}
\usepackage{lipsum}
\usepackage{mathtools}
\usepackage{cuted}
\usepackage{braket}
\usepackage{booktabs}
\usepackage{mdframed}

\DeclareMathOperator{\tr}{Tr}
\DeclareMathOperator{\vc}{vec}

\newcommand{\cutt}[1]{}
\newcommand{\cut}[1]{}
\usepackage{multirow}

\newtheorem{definition}[]{Definition}

\usepackage{subcaption}

\title{Unrolling SVT to obtain computationally efficient SVT for n-qubit quantum state tomography}

\author{ Siva Shanmugam* \thanks{*The authors are with the Department of Electrical Engineering, Indian Institute of Technology Madras, Chennai, India 600 036. Email: \texttt{ee17s024@smail.iitm.ac.in,skalyani@ee.iitm.ac.in}} \hspace{16pt} Sheetal Kalyani*
}

\pgfplotsset{compat=1.17}
\begin{document}
	\maketitle
    \begin{abstract}
Quantum state tomography aims to estimate the state of a quantum mechanical system which is described by a trace one, Hermitian positive semidefinite complex matrix, given a set of measurements of the state. Existing works focus on estimating the density matrix that represents the state, using a compressive sensing approach, with only fewer measurements than that required for a tomographically complete set, with the assumption that the true state has a low rank. One very popular method to estimate the state is the use of the Singular Value Thresholding (SVT) algorithm. In this work, we present a machine learning approach to estimate the quantum state of n-qubit systems by unrolling the iterations of SVT which we call Learned Quantum State Tomography (LQST). As merely unrolling SVT may not ensure that the output of the network meets the constraints required for a quantum state,  we design and train a custom neural network whose architecture is inspired from the iterations of SVT with additional layers to meet the required constraints. We show that our proposed LQST with very few layers reconstructs the density matrix with much better fidelity than the SVT algorithm which takes many hundreds of iterations to converge. We also demonstrate the reconstruction of the quantum Bell state from an informationally incomplete set of noisy measurements.
\end{abstract}

\section{Introduction} \label{sec:intro}
Quantum state tomography plays a key role in building computers that exploit the laws of quantum mechanics \cite{qcqi}. For instance, it is used to verify that a certain source of quantum system outputs the systems in the desired quantum state. It does so by measuring the output systems from the source and reconstructing the underlying state that the systems are prepared in.

Mathematically, quantum state tomography of finite quantum mechanical systems is the problem of reconstructing unit trace, positive semidefinite complex matrix called the density matrix from a set of measurements that are linear functions of the matrix. If an informationally complete set of measurements are available then one of the first approaches is to invert the linear relation between the state and the measurements to obtain the desired estimate for the state \cite{onthemeasofqubits}. This approach gives a closed-form expression to the estimate but suffers from two major problems. 1) It requires an informationally complete set of measurements. The number of measurements in this set is exponential in the size of the system. 2) The estimated state will not necessarily be positive semidefinite. Another approach is to define a likelihood function for the measurements and estimate the state by maximising the likelihood for the observed measurements. \cite{mlmethodsinqst} and \cite{mlqst} discuss the maximum likelihood (ML) methods in detail. In this approach, the density matrix is parameterized to ensure positive semidefiniteness and the best parameters that maximize the likelihood is found \cite{mlestofdensitymatrix},\cite{dilutedml}. This approach again requires a large set of measurements to get a better estimate. \cite{bayesianinference} assumes a prior distribution to the above said parameters and the mean of the posterior distribution is used to reconstruct the state. However, finding the mean of the posterior demands Markov Chain Monte Carlo (MCMC) algorithm which is computationally complex to implement. 

Recently, the problem of estimating low-rank matrices using a compressive sensing approach has drawn a lot of attention. The work in \cite{lowrankanybasisgross} quantified the number of measurements required for the success of matrix recovery from its linear measurements. In \cite{psdmf} positive semidefinite matrix factorization (PSDMF) is used to estimate low-rank matrices. In \cite{lrmscaledsubgradient} a subgradient method is performed to estimate the matrix. In \cite{powerfactorization} authors proposed the power factorization algorithm to estimate low rank matrices and the idea was further extended in \cite{als} and \cite{sparesepowerfactorization}. The estimation of the positive semidefinite matrix from its linear measurements has been analyzed in \cite{rankonemeas} and \cite{psdrlm}. Following these, the focus in quantum estimation is mostly on using compressive sensing approach to estimate quantum state with low rank when only few measurements are available \cite{permutationallyinvariant}, \cite{qstsingleobservable}, \cite{gross}, \cite{gross2}, \cite{qtprotocols}. This approach demands only a few measurements to estimate the low-rank state. While, \cite{gross} uses rank minimization method introduced in \cite{svt} to recover the low rank quantum state. The work in \cite{gross2} introduces a least-squares method for quantum estimation.

The capability of computing machines to store and process a huge amount of data and the capability of neural networks to learn very complicated functions have enabled neural networks to find their roots in almost all fields including quantum state tomography. For instance, in \cite{rqst_gm} the authors have used RBM (Restricted Boltzmann Machine) to learn quantum states, and the same was analyzed with real-time data in \cite{nnqst2qubit}. However, it has been very hard to interpret the deep neural networks. In this context, unrolled algorithms which are deep neural networks whose architecture is inspired from interpretable classical algorithms have attracted considerable attention recently \cite{unrollingsurvey}. To this end, we introduce a machine learning approach to estimate the state of the n-qubit system by designing and training a custom neural network inspired from SVT \cite{gross} that outperforms SVT and also has significantly lesser computational complexity compared with SVT. Our network can estimate low-rank quantum states from an informationally incomplete set of measurements with high accuracy. We test our network by numerically estimating random density matrices and 2-qubit quantum Bell state from an informationally incomplete set of measurements.

\subsection{Our contributions and outline of work}

We design and train a custom neural network whose layers are inspired from the iterations of SVT and a few more layers are designed at the end to ensure that the output of the network meets the constraints required for a quantum state. These final layers essentially perform the projection operations which project the output from the previous layers to the required constraint set. We numerically simulate the quantum state tomography of 4-qubit system and compare it with the SVT algorithm. We term our proposed method Learned Quantum State Tomography (LQST). The advantages of the proposed method are as follows. Firstly, our proposed LQST outperforms SVT in terms of fidelity between the true and estimated states when the parameters of SVT are tuned for faster convergence. Secondly, the computational complexity of our LQST is much less compared with that of SVT. A 3-layered LQST outperforms SVT which converges in 1700 iterations. If the parameters of SVT are tuned for better performance (in terms of fidelity) as opposed to tuning for faster convergence, SVT takes many thousands of iterations to converge whereas our LQST achieves that same fidelity in three or four layers. Thirdly, LQST directly reconstructs the density matrix of a general quantum state from an informationally incomplete set of measurements whereas another neural network-based quantum estimation which uses Restricted Boltzmann Machine (RBM) \cite{rqst_gm} learns only the measurement probabilities of a particular quantum state.

\subsection{Background}
The state of a $k$-dimensional quantum mechanical system is completely described by a density matrix $\rho \in \mathbb{C}^{k \times k}$ which is Hermitian, positive semidefinite and has unit trace \cite{qcqi}. Unlike the classical systems, when a certain observable for the quantum mechanical system is measured, the measurement operation probabilistically yields one of $k$ possible outcomes, where the probabilities of the outcomes depend on the state in which the system is. Further, after the measurement operation, the system goes to a new state depending on the outcome. Since the state of the system changes post-measurement, a new copy of the system in the same state as the previous one is required to make a new measurement. This can be thought of as follows. Consider a source of quantum mechanical system that outputs the quantum systems in a particular state. Every time the source outputs a system, an observable for the system is measured. These measurements are then used to reconstruct the state that the source is emitting the systems in. These observables are represented by Hermitian matrices. The Pauli observables which we use in this work are explicitly given in section \ref{sec:problem}. In this work, we focus on estimating the state of $n$-qubit systems which are $d$-dimensional systems with $d=2^n$ by measuring the Pauli observables. 

\subsection{Notations used}
We use lower case letters for scalars. Vectors are denoted by bold lower case letters such as $\bm b$ and the $i^{th}$ element of $\bm b$ is denoted as $b_i$. Upper case letters are used for denoting matrices and sets. Hermitian of the matrix $A$ is denoted as $A^\dagger$. The diagonal matrix with diagonal elements $a_1$ through $a_m$ is denoted as diag$(a_1 \dots a_m)$. The square root matrix of a positive definite matrix is denoted as $\sqrt{A}$. $|A|$ denotes the square root of the positive definite matrix $A^\dagger A$. The tensor product between two matrices, say $A$ and $B$ is represented as $A \otimes B$.  $\vc(A)$ represents the vector obtained by vectorizing $A$. $\norm{.}_2$ denotes the euclidean norm. $\norm{A}_F$ denotes the Frobenius norm of the matrix $A$. $\norm{A}_{*}$ represents the nuclear norm of $A$ which is the sum of absolute singular values of $A$.

\subsection{Organisation}

In the next section, the problem is formulated and the figures of merit that we use to quantify the performance of estimators are defined. In section \ref{sec:svt}, the existing compressive sensing approach Singular Value Thresholding (SVT) for low-rank state estimation is discussed. Our machine learning approach to quantum state tomography is presented in the section \ref{sec:LQST}. We present the numerical simulations in section \ref{sec:simulation} and conclude in section \ref{sec:conclusion}.
    \section{Problem Formulation} \label{sec:problem}

Let $\rho \in \mathbb{C}^{d \times d}$ be a rank-$r$, positive semidefinite, unit trace density matrix that represents the state of $n$-qubit quantum system with $d=2^n$. Before defining the measurements of the state, the Pauli operators and the identity operator on a single qubit system are defined in Table \ref{tab:pauliandidentity}. 

\begin{table}[h]
\centering
\begin{tabular}{@{}|l|l|l|l|@{}}
\toprule
$X_1$ & $X_2$ & $X_3$ & $X_4$ \\ \midrule
$\begin{bmatrix} 0 &1 \\ 1 &0 \end{bmatrix}$ & $\begin{bmatrix} 0 &-i \\ i &0 \end{bmatrix}$ & $\begin{bmatrix} 1 &0 \\ 0 &-1 \end{bmatrix}$ & $\begin{bmatrix} 1 &0 \\ 0 &1 \end{bmatrix}$ \\ \bottomrule
\end{tabular}
\caption{Pauli operators $(X_1, X_2, X_3)$ and the identity operator on a single qubit system $(X_4).$}
\label{tab:pauliandidentity}
\end{table}

The Pauli operators on the $n$-qubit system is given by $\{A_i=X_{i_1} \otimes \dots \otimes X_{i_n} \in \mathbb{C}^{d \times d}\}_{i=1}^{d^2}$, where $i_1, \dots i_n \in \{1,2,3,4\}$. These Pauli operators are also called the observables in the quantum literature. These observables can be experimentally measured for a quantum mechanical system in a particular state. Measuring an observable $A_i$ will probabilistically yield one of $d$ outcomes where the probabilities of the outcomes depend on the state of the system $\rho$. The expected outcome when an observable, say $A_i$, for the system in state $\rho$ is measured is given by
\begin{equation}
    \langle A_i \rangle := \tr[A_i\rho]
\end{equation} These expected values are real numbers and can be experimentally obtained by measuring the observables repeatedly and finding the average of the outcomes.  Since $A_{d^2}$ is the identity operator and the density matrix has unit trace, $\langle A_{d^2} \rangle$ is always unity. The set of observables $\mathcal{O} := \{A_1, \dots A_{d^2-1}\}$ represents a tomographically complete set of measurements, in the sense that measuring these observables will yield complete information about the state of the system. In what follows, measurements mean these expected values. The $d^2-1$ measurements $\{\langle A_i \rangle\}_{i=1}^{d^2-1}$ completely describe the state of the system. 

However, for the $n$-qubit system, the dimension of the density matrix $d$ scales exponentially with $n$ as $d=2^n$. Because of this, it is very tedious to obtain the entire $d^2-1$ measurements. Hence we focus on recovering the state of the system with fewer measurements, say $m < d^2-1$.

Given $m$ observables $\{A_{j_1}, A_{j_2} \dots A_{j_m}\} \subseteq \mathcal{O} $ where $j_1, \dots j_{m-1} \in \{1, 2, \dots d^2-1\}$, the $m$ measurements of the state $\rho$ is modelled as
\begin{equation}
    \bm{b} = \mathcal{A}(\rho) \label{eqn:qmes}
\end{equation}where $\bm b, \mathcal{A}(\rho)$ and the linear map $\mathcal{A}: \mathbb{C}^{d \times d}\to \mathbb{C}^m$ is defined as
\begin{align}
    \mathcal{A}_i(X)=
    \tr[A_{j_i}X], \quad &\text{for }i=1,\dots m
    \label{eqn:qmap}
\end{align} Though the map $\mathcal{A}$ is defined from $\mathbb{C}^{d \times d}$ to $\mathbb{C}^m, \mathcal{A}(\rho)$ is always a real vector. The map is, however, defined between two complex vector spaces to help the computation of it's adjoint.  Given these $m$ Pauli observables and $m$ measurements, our goal is to recover the density matrix $\rho$. 
As the figures of merit we use trace distance and fidelity as defined below. 
\begin{definition}Trace distance between the two density matrices $\rho$ and $\sigma$ is defined as 
\begin{equation}
    \mathcal{D}(\rho,\sigma) = \frac{1}{2} \tr[|\rho - \sigma|]
\end{equation}
\end{definition}
\begin{definition}
Fidelity between two density matrices $\rho$ and $\sigma$ is defined as
\begin{equation}
    \mathcal{F}(\rho - \sigma) = \tr\left[\sqrt{\sqrt{\rho}\sigma\sqrt{\rho}}\right]
\end{equation}
\end{definition}
Trace distance can also be expressed as 0.5 times the sum of singular values of the difference matrix $\rho - \sigma$ which is lower when the states are closer. However, fidelity is higher when the states are closer. It is bounded between 0 and 1 and is equal to 1 if and only if the states are same i.e., $\rho = \sigma$. 

We first discuss the existing SVT algorithm in the next section and then present our neural network to perform quantum state tomography.

    \section{SVT algorithm} \label{sec:svt}

The SVT algorithm was introduced in \cite{svt} to recover real matrices from general measurements such as convex functions of the matrix. We discuss in this section, a special case of the SVT algorithm concerning the recovery of real matrices from its linear measurements. SVT solves affine rank minimization formulated in \eqref{eqn:arm} as
\begin{subequations}
    \begin{alignat}{2}
    &\!\min_{X \in \mathbb{R}^{d \times d}}        &\qquad& \text{Rank}(X)  \\
    &\text{such that} &      & \mathcal{A}(X)=\bm b
    \end{alignat}  \label{eqn:arm}
\end{subequations}
by minimizing the nuclear norm of the matrix which is a surrogate to the rank function. Specifically, the convex optimization problem that the SVT solves is given as \cite{svt}
\begin{subequations}
    \begin{alignat}{2}
    &\!\min_{X \in \mathbb{R}^{d \times d}}        &\qquad& \tau \norm{X}_{*} + \frac{1}{2} \norm{X}_F^2   \\
    &\text{such that} &      & \mathcal{A}(X)=\bm b
    \end{alignat}  \label{eqn:svtproblem}
\end{subequations} where $\tau>0$ is a constant and $\mathcal{A}:\mathcal{R}^{d \times d}\to \mathcal{R}^m$ is a linear map. In theorem 3.1 of \cite{svt}, authors proved that in the limit $\tau$ tending to infinity, the solutions to \eqref{eqn:svtproblem} converge to the matrix with minimum trace norm that is also consistent with the measurements i.e., ($\mathcal{A}(X) = \bm b$). In \cite{arm} authors proved that minimizing the nuclear norm yields the minimum rank solution with high probability (see theorem 3.3 and theorem 4.2 of \cite{arm}). Hence SVT minimizes rank formulated in \eqref{eqn:arm} by solving \eqref{eqn:svtproblem}. 

Note that the problem \eqref{eqn:svtproblem} is a constrained convex optimization problem and the strong duality holds for the problem \cite{boyd}. The Lagrangian function for the problem is given as
\begin{equation}
    \mathcal{L}(X, \bm y) = \tau \norm{X}_{*} + \frac{1}{2} \norm{X}_F^2 + \bm y^T (\bm b - \mathcal{A}(X))
\end{equation}
where $X$ is the optimization variable and $\bm y$ is the Lagrangian variable. Since strong duality holds for the problem, finding the saddle point $(X^*, \bm y^*)$ of the Lagrangian $\mathcal{L}(X, \bm y)$ gives the solution ($X^*$) to \eqref{eqn:svtproblem}. The saddle point is written as
\begin{equation}
    \underset{\bm y}{\sup} \, \underset{X}{\inf} \, \mathcal{L}(X, \bm y) = \mathcal{L}(X^*,\bm y^*) = \underset{X}{\inf} \, \underset{\bm y}{\sup} \, \mathcal{L}(X, \bm y)
\end{equation}

Authors in \cite{svt} used Uzawa's iterations \cite{uzawa} to find the saddle point of the Lagrangian. Uzawa's iterations starts with an initial $\bm y^0$ and repeats two steps until convergence. In the first step, the minimizer (say $X^k$) of $\mathcal{L}(., \bm y)$ for the given $\bm y$ is found. In the second step, a gradient ascent step is taken along the direction $\bm y$ for the given $X^k$ found in the previous step. These steps are given as follows
\begin{align}
    \begin{cases}
    X^k &= \underset{X}{\arg \min} \, \mathcal{L}(X, \bm y^{k-1}) \\
    \bm y^k &= \bm y^{k-1} + \delta_k \nabla_{\bm y} \mathcal{L}(X^k , \bm y)
    \end{cases} \label{eqn:uzawaiterations}
\end{align} where $\delta_k > 0$ is the step size used for gradient ascent in the $k^{th}$ iteration. The gradient $\nabla_{\bm y} \mathcal{L}(X^k , \bm y)$ is given as $\mathcal{A}(X^k) - \bm b$. The closed form solution to the minimization problem in \eqref{eqn:uzawaiterations} is given as
\begin{equation}
    \underset{X}{\arg \min} \, \mathcal{L}(X, \bm y^{k-1}) = \mathcal{D}_{\tau}( \mathcal{A}^*(\bm y^{k-1})) \label{eqn:minalongX}
\end{equation}
where the operator $\mathcal{A}^*:\mathbb{R}^m \to \mathbb{R}^{d \times d}$ is the adjoint of the linear map $\mathcal{A}$ and is defined as
\begin{equation}
    \mathcal{A}^*(\bm y) = \sum_{i=1}^m y_i A_i^T \label{eqn:adjoint}
\end{equation} and the operator $\mathcal{D}_\tau:\mathbb{R}^{d \times d} \to \mathbb{R}^{d \times d}$ is the singular value thresholding operator and is defined as
\begin{equation}
    \mathcal{D}_{\tau}(X) = U \mathcal{D}_{\tau}(\Sigma)V^T \label{eqn:dtau}
\end{equation}
where the singular value decomposition of $X$ is given as $X = U \Sigma V^T$, $\Sigma$ is the diagonal matrix with the singular values of $X$ in its diagonal positions which we write as $\Sigma = \text{diag}(\sigma_1(X), \dots \sigma_r(X))$ where $r$ represents the rank of $X$. $\mathcal{D}_{\tau}(\Sigma)$ is a diagonal matrix whose non zero entries are found by soft thresolding the entries of $\Sigma$ and is given as $\mathcal{D}_{\tau}(\Sigma) = \text{diag}(\max(\sigma_1(X) - \tau, 0), \dots \max(\sigma_r(X) - \tau, 0))$. 
Using the gradient of $\mathcal{L}$ along $\bm y$ and the equations \eqref{eqn:minalongX}, \eqref{eqn:adjoint} and \eqref{eqn:dtau} Uzawa's iterations \eqref{eqn:uzawaiterations} can be rewritten as
\begin{align}
    \begin{cases}
    X^k &= \mathcal{D}_{\tau}( \mathcal{A}^*(\bm y^{k-1}))\\
    \bm y^k &= \bm y^{k-1} + \delta_k 
    (\bm b - \mathcal{A}(X^k))
    \end{cases} \label{eqn:uzawaiterationsfinal}
\end{align}
where $\delta_k > 0$ is the stepsize to do gradient ascent, $\tau > 0$ is the threshold value used in singular value thresholding operator and the maps $\mathcal{A}$ and $\mathcal{A}^*$ are defined using the measurement matrices $A_1$ through $A_m$ and \eqref{eqn:adjoint}. 

SVT algorithm repeatedly performs the steps in \eqref{eqn:uzawaiterationsfinal} which requires the knowledge of the right choice of $\delta_k$ and $\tau$ to solve the affine rank minimization problem \eqref{eqn:arm}. In the subsequent sections, we design a trainable deep network based on the steps in \eqref{eqn:uzawaiterationsfinal} and make necessary modifications to estimate the quantum state which is robust to the parameters such as $\delta$ and $\tau$.

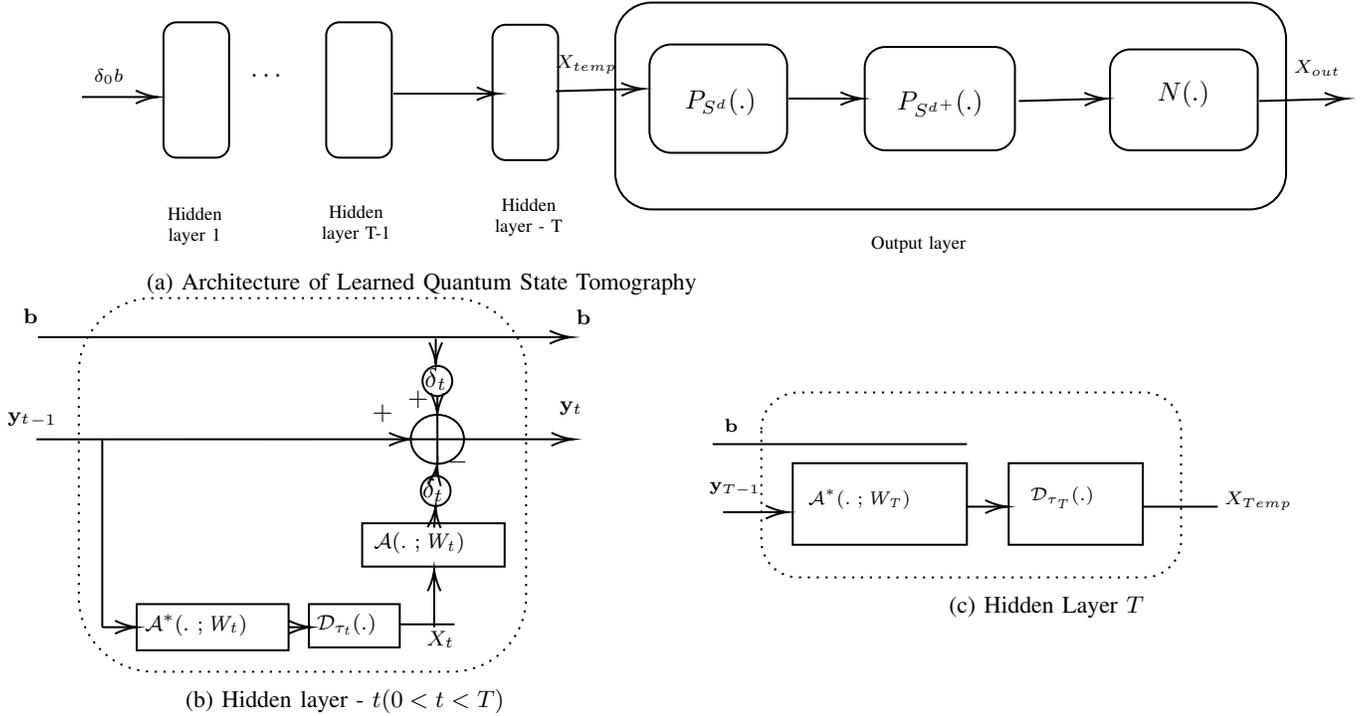
\begin{figure*}[ht!]
\hspace{9mm}
\begin{subfigure}[H]{.5\textwidth}
\centering

\tikzset{every picture/.style={line width=0.75pt}} 

\begin{tikzpicture}[x=0.75pt,y=0.75pt,yscale=-1,xscale=1]

\draw   (56.07,57.56) .. controls (56.07,53.9) and (59.03,50.94) .. (62.68,50.94) -- (82.53,50.94) .. controls (86.18,50.94) and (89.15,53.9) .. (89.15,57.56) -- (89.15,112.63) .. controls (89.15,116.28) and (86.18,119.24) .. (82.53,119.24) -- (62.68,119.24) .. controls (59.03,119.24) and (56.07,116.28) .. (56.07,112.63) -- cycle ;
\draw    (14.94,88.63) -- (51.02,88.63) ;
\draw [shift={(53.02,88.63)}, rotate = 180] [color={rgb, 255:red, 0; green, 0; blue, 0 }  ][line width=0.75]    (10.93,-3.29) .. controls (6.95,-1.4) and (3.31,-0.3) .. (0,0) .. controls (3.31,0.3) and (6.95,1.4) .. (10.93,3.29)   ;
\draw    (253.93,85.69) -- (295.96,84.54) ;
\draw [shift={(297.96,84.49)}, rotate = 538.4300000000001] [color={rgb, 255:red, 0; green, 0; blue, 0 }  ][line width=0.75]    (10.93,-3.29) .. controls (6.95,-1.4) and (3.31,-0.3) .. (0,0) .. controls (3.31,0.3) and (6.95,1.4) .. (10.93,3.29)   ;
\draw    (172.04,86.94) -- (223,86.94) ;
\draw [shift={(225,86.94)}, rotate = 180] [color={rgb, 255:red, 0; green, 0; blue, 0 }  ][line width=0.75]    (10.93,-3.29) .. controls (6.95,-1.4) and (3.31,-0.3) .. (0,0) .. controls (3.31,0.3) and (6.95,1.4) .. (10.93,3.29)   ;
\draw   (138.13,57.56) .. controls (138.13,53.9) and (141.09,50.94) .. (144.75,50.94) -- (164.59,50.94) .. controls (168.25,50.94) and (171.21,53.9) .. (171.21,57.56) -- (171.21,112.63) .. controls (171.21,116.28) and (168.25,119.24) .. (164.59,119.24) -- (144.75,119.24) .. controls (141.09,119.24) and (138.13,116.28) .. (138.13,112.63) -- cycle ;
\draw   (222.02,58.8) .. controls (222.02,55.14) and (224.98,52.18) .. (228.64,52.18) -- (248.48,52.18) .. controls (252.14,52.18) and (255.1,55.14) .. (255.1,58.8) -- (255.1,113.87) .. controls (255.1,117.52) and (252.14,120.48) .. (248.48,120.48) -- (228.64,120.48) .. controls (224.98,120.48) and (222.02,117.52) .. (222.02,113.87) -- cycle ;
\draw   (301.17,73.05) .. controls (301.17,67.02) and (306.06,62.12) .. (312.1,62.12) -- (359.67,62.12) .. controls (365.7,62.12) and (370.6,67.02) .. (370.6,73.05) -- (370.6,105.85) .. controls (370.6,111.89) and (365.7,116.79) .. (359.67,116.79) -- (312.1,116.79) .. controls (306.06,116.79) and (301.17,111.89) .. (301.17,105.85) -- cycle ;
\draw    (371.34,89.42) -- (406.47,89.45) ;
\draw [shift={(408.47,89.45)}, rotate = 180.05] [color={rgb, 255:red, 0; green, 0; blue, 0 }  ][line width=0.75]    (10.93,-3.29) .. controls (6.95,-1.4) and (3.31,-0.3) .. (0,0) .. controls (3.31,0.3) and (6.95,1.4) .. (10.93,3.29)   ;
\draw   (409.73,73.8) .. controls (409.73,68.04) and (414.41,63.36) .. (420.17,63.36) -- (475.04,63.36) .. controls (480.81,63.36) and (485.48,68.04) .. (485.48,73.8) -- (485.48,105.11) .. controls (485.48,110.87) and (480.81,115.55) .. (475.04,115.55) -- (420.17,115.55) .. controls (414.41,115.55) and (409.73,110.87) .. (409.73,105.11) -- cycle ;
\draw    (487.48,90.66) -- (529.51,89.51) ;
\draw [shift={(531.51,89.45)}, rotate = 538.4300000000001] [color={rgb, 255:red, 0; green, 0; blue, 0 }  ][line width=0.75]    (10.93,-3.29) .. controls (6.95,-1.4) and (3.31,-0.3) .. (0,0) .. controls (3.31,0.3) and (6.95,1.4) .. (10.93,3.29)   ;
\draw   (533.45,74.79) .. controls (533.45,69.17) and (538.01,64.61) .. (543.64,64.61) -- (597.75,64.61) .. controls (603.37,64.61) and (607.94,69.17) .. (607.94,74.79) -- (607.94,105.36) .. controls (607.94,110.98) and (603.37,115.55) .. (597.75,115.55) -- (543.64,115.55) .. controls (538.01,115.55) and (533.45,110.98) .. (533.45,105.36) -- cycle ;
\draw    (608.67,90.66) -- (650.7,89.51) ;
\draw [shift={(652.7,89.45)}, rotate = 538.4300000000001] [color={rgb, 255:red, 0; green, 0; blue, 0 }  ][line width=0.75]    (10.93,-3.29) .. controls (6.95,-1.4) and (3.31,-0.3) .. (0,0) .. controls (3.31,0.3) and (6.95,1.4) .. (10.93,3.29)   ;
\draw   (283.91,59.26) .. controls (283.91,49.18) and (292.09,41) .. (302.18,41) -- (603.98,41) .. controls (614.07,41) and (622.24,49.18) .. (622.24,59.26) -- (622.24,127.1) .. controls (622.24,137.19) and (614.07,145.36) .. (603.98,145.36) -- (302.18,145.36) .. controls (292.09,145.36) and (283.91,137.19) .. (283.91,127.1) -- cycle ;

\draw (52.17,142.54) node [anchor=north west][inner sep=0.75pt]  [font=\scriptsize] [align=left] {\begin{minipage}[lt]{27.34pt}\setlength\topsep{0pt}
\begin{center}
Hidden \\layer 1
\end{center}

\end{minipage}};
\draw (131.54,141.63) node [anchor=north west][inner sep=0.75pt]  [font=\scriptsize] [align=left] {\begin{minipage}[lt]{30.38pt}\setlength\topsep{0pt}
\begin{center}
Hidden \\layer T-1
\end{center}

\end{minipage}};
\draw (31.61,117.26) node [anchor=north west][inner sep=0.75pt]   [align=left] {$ $};
\draw (19.59,72.69) node [anchor=north west][inner sep=0.75pt]  [font=\scriptsize] [align=left] {$\displaystyle \delta_0 b$};
\draw (252.46,66.24) node [anchor=north west][inner sep=0.75pt]  [font=\scriptsize] [align=left] {$\displaystyle X_{temp}$};
\draw (219.65,138.03) node [anchor=north west][inner sep=0.75pt]  [font=\scriptsize] [align=left] {\begin{minipage}[lt]{28.78pt}\setlength\topsep{0pt}
\begin{center}
Hidden \\layer - T
\end{center}

\end{minipage}};
\draw (98.53,75.49) node [anchor=north west][inner sep=0.75pt]   [align=left] {$\displaystyle \dotsc $};
\draw (317.59,82.94) node [anchor=north west][inner sep=0.75pt]   [align=left] {$\displaystyle P_{S^{d}}( .)$};
\draw (424.55,83.3) node [anchor=north west][inner sep=0.75pt]   [align=left] {$\displaystyle P_{S{^{d}}^{+}}( .)$};
\draw (555.85,78.76) node [anchor=north west][inner sep=0.75pt]   [align=left] {$\displaystyle N( .)$};
\draw (407.72,157.45) node [anchor=north west][inner sep=0.75pt]  [font=\scriptsize] [align=left] {\begin{minipage}[lt]{41.62pt}\setlength\topsep{0pt}
\begin{center}
Output layer
\end{center}

\end{minipage}};
\draw (624.82,68.73) node [anchor=north west][inner sep=0.75pt]  [font=\scriptsize] [align=left] {$\displaystyle X_{out}$};
\end{tikzpicture}
\caption{Architecture of Learned Quantum State Tomography}
\label{fig:architechturelqst}
\end{subfigure} \\
\hspace{12mm}
\begin{subfigure}[H]{.5\textwidth}
\tikzset{every picture/.style={line width=0.75pt}} 

\begin{tikzpicture}[x=0.75pt,y=0.75pt,yscale=-1,xscale=1]

\draw    (202.97,205.6) -- (218.34,205.6) ;
\draw [shift={(220.34,205.6)}, rotate = 180] [color={rgb, 255:red, 0; green, 0; blue, 0 }  ][line width=0.75]    (10.93,-3.29) .. controls (6.95,-1.4) and (3.31,-0.3) .. (0,0) .. controls (3.31,0.3) and (6.95,1.4) .. (10.93,3.29)   ;
\draw    (370.37,205.6) -- (370.37,178.46) ;
\draw [shift={(370.37,176.46)}, rotate = 450] [color={rgb, 255:red, 0; green, 0; blue, 0 }  ][line width=0.75]    (10.93,-3.29) .. controls (6.95,-1.4) and (3.31,-0.3) .. (0,0) .. controls (3.31,0.3) and (6.95,1.4) .. (10.93,3.29)   ;
\draw   (334.05,153.15) -- (405.91,153.15) -- (405.91,175.01) -- (334.05,175.01) -- cycle ;
\draw    (202.97,110.41) -- (202.97,205.6) ;
\draw   (358.56,110.74) .. controls (358.56,103.9) and (364.57,98.36) .. (371.98,98.36) .. controls (379.4,98.36) and (385.41,103.9) .. (385.41,110.74) .. controls (385.41,117.58) and (379.4,123.13) .. (371.98,123.13) .. controls (364.57,123.13) and (358.56,117.58) .. (358.56,110.74) -- cycle ; \draw   (358.56,110.74) -- (385.41,110.74) ; \draw   (371.98,98.36) -- (371.98,123.13) ;
\draw    (371.16,60.02) -- (371.16,70.84) ;
\draw [shift={(371.16,72.84)}, rotate = 270] [color={rgb, 255:red, 0; green, 0; blue, 0 }  ][line width=0.75]    (10.93,-3.29) .. controls (6.95,-1.4) and (3.31,-0.3) .. (0,0) .. controls (3.31,0.3) and (6.95,1.4) .. (10.93,3.29)   ;
\draw    (370.37,155.34) -- (371,146.44) ;
\draw [shift={(371.14,144.44)}, rotate = 454.02] [color={rgb, 255:red, 0; green, 0; blue, 0 }  ][line width=0.75]    (10.93,-3.29) .. controls (6.95,-1.4) and (3.31,-0.3) .. (0,0) .. controls (3.31,0.3) and (6.95,1.4) .. (10.93,3.29)   ;
\draw    (353,204.15) -- (380.64,204.15) ;
\draw  [dash pattern={on 0.84pt off 2.51pt}] (191.13,77.2) .. controls (191.13,56.37) and (208.01,39.49) .. (228.85,39.49) -- (379.03,39.49) .. controls (399.86,39.49) and (416.74,56.37) .. (416.74,77.2) -- (416.74,190.36) .. controls (416.74,211.19) and (399.86,228.08) .. (379.03,228.08) -- (228.85,228.08) .. controls (208.01,228.08) and (191.13,211.19) .. (191.13,190.36) -- cycle ;
\draw   (307.2,193.95) -- (353,193.95) -- (353,215.8) -- (307.2,215.8) -- cycle ;
\draw   (220.34,193.95) -- (296.94,193.95) -- (296.94,217.24) -- (220.34,217.24) -- cycle ;
\draw    (296.94,205.6) -- (305.99,205.6) ;
\draw [shift={(307.99,205.6)}, rotate = 180] [color={rgb, 255:red, 0; green, 0; blue, 0 }  ][line width=0.75]    (10.93,-3.29) .. controls (6.95,-1.4) and (3.31,-0.3) .. (0,0) .. controls (3.31,0.3) and (6.95,1.4) .. (10.93,3.29)   ;
\draw    (169.78,110.77) -- (356.56,110.74) ;
\draw [shift={(358.56,110.74)}, rotate = 539.99] [color={rgb, 255:red, 0; green, 0; blue, 0 }  ][line width=0.75]    (10.93,-3.29) .. controls (6.95,-1.4) and (3.31,-0.3) .. (0,0) .. controls (3.31,0.3) and (6.95,1.4) .. (10.93,3.29)   ;
\draw    (385.41,110.74) -- (438.44,110.74) ;
\draw [shift={(440.44,110.74)}, rotate = 180] [color={rgb, 255:red, 0; green, 0; blue, 0 }  ][line width=0.75]    (10.93,-3.29) .. controls (6.95,-1.4) and (3.31,-0.3) .. (0,0) .. controls (3.31,0.3) and (6.95,1.4) .. (10.93,3.29)   ;
\draw    (170.59,59.23) -- (437.56,59.23) ;
\draw [shift={(439.56,59.23)}, rotate = 180] [color={rgb, 255:red, 0; green, 0; blue, 0 }  ][line width=0.75]    (10.93,-3.29) .. controls (6.95,-1.4) and (3.31,-0.3) .. (0,0) .. controls (3.31,0.3) and (6.95,1.4) .. (10.93,3.29)   ;
\draw   (364.48,81.31) .. controls (364.48,77.11) and (367.84,73.7) .. (371.98,73.7) .. controls (376.13,73.7) and (379.48,77.11) .. (379.48,81.31) .. controls (379.48,85.52) and (376.13,88.93) .. (371.98,88.93) .. controls (367.84,88.93) and (364.48,85.52) .. (364.48,81.31) -- cycle ;
\draw    (371.98,88.93) -- (371.98,96.36) ;
\draw [shift={(371.98,98.36)}, rotate = 270] [color={rgb, 255:red, 0; green, 0; blue, 0 }  ][line width=0.75]    (10.93,-3.29) .. controls (6.95,-1.4) and (3.31,-0.3) .. (0,0) .. controls (3.31,0.3) and (6.95,1.4) .. (10.93,3.29)   ;
\draw   (363.64,136.82) .. controls (363.64,132.62) and (367,129.21) .. (371.14,129.21) .. controls (375.28,129.21) and (378.64,132.62) .. (378.64,136.82) .. controls (378.64,141.03) and (375.28,144.44) .. (371.14,144.44) .. controls (367,144.44) and (363.64,141.03) .. (363.64,136.82) -- cycle ;
\draw    (371.14,129.21) -- (371.71,125.11) ;
\draw [shift={(371.98,123.13)}, rotate = 457.92] [color={rgb, 255:red, 0; green, 0; blue, 0 }  ][line width=0.75]    (10.93,-3.29) .. controls (6.95,-1.4) and (3.31,-0.3) .. (0,0) .. controls (3.31,0.3) and (6.95,1.4) .. (10.93,3.29)   ;

\draw (338.54,156.22) node [anchor=north west][inner sep=0.75pt]  [font=\footnotesize] [align=left] {$\displaystyle \mathcal{A}( .\ ;\mathit{W_t}) \ $};
\draw (355.6,84.63) node [anchor=north west][inner sep=0.75pt]   [align=left] {$\displaystyle +$};
\draw (375.52,116.25) node [anchor=north west][inner sep=0.75pt]   [align=left] {$\displaystyle -$};
\draw (337.7,91.39) node [anchor=north west][inner sep=0.75pt]   [align=left] {$\displaystyle +$};
\draw (161.76,43.21) node [anchor=north west][inner sep=0.75pt]  [font=\footnotesize] [align=left] {$\displaystyle \mathbf{b}$};
\draw (153.88,95.51) node [anchor=north west][inner sep=0.75pt]  [font=\footnotesize] [align=left] {$\displaystyle \mathbf{y}_{t}{}_{-1}$};
\draw (432.2,90.38) node [anchor=north west][inner sep=0.75pt]  [font=\footnotesize] [align=left] {$\displaystyle \mathbf{y}_{t}{}$};
\draw (365.18,205.95) node [anchor=north west][inner sep=0.75pt]  [font=\footnotesize] [align=left] {$\displaystyle X_{t}$};
\draw (309.74,197.45) node [anchor=north west][inner sep=0.75pt]  [font=\footnotesize] [align=left] {$\displaystyle \mathcal{D}_{\mathit{\tau_t}}( .) \ $};
\draw (332.22,195.88) node [anchor=north west][inner sep=0.75pt]   [align=left] {$ $};
\draw (223.15,197.45) node [anchor=north west][inner sep=0.75pt]  [font=\footnotesize] [align=left] {$\displaystyle \mathcal{A}^{*}( .\ ;\mathit{W_t}) \ $};
\draw (440.66,43.87) node [anchor=north west][inner sep=0.75pt]  [font=\footnotesize] [align=left] {$\displaystyle \mathbf{b}$};
\draw (364.2,73.4) node [anchor=north west][inner sep=0.75pt]   [align=left] {$\mathit{\delta_t}$};

\draw (363,129.5) node [anchor=north west][inner sep=0.75pt]   [align=left] {${\delta_t}$};

\end{tikzpicture}

    \caption{Hidden layer - $t  (0 < t < T)$}
    \label{fig:hiddenlayerTm1}
\end{subfigure}
\hspace{1mm}
\begin{subfigure}[H]{.5\textwidth}

\tikzset{every picture/.style={line width=0.75pt}} 

\begin{tikzpicture}[x=0.75pt,y=0.75pt,yscale=-1,xscale=1]

\draw   (311.3,81.7) -- (378.93,81.7) -- (378.93,122.88) -- (311.3,122.88) -- cycle ;
\draw   (202.37,81.7) -- (290.22,81.7) -- (290.22,122.88) -- (202.37,122.88) -- cycle ;
\draw    (167.24,106.41) -- (197.74,106.41) ;
\draw [shift={(199.74,106.41)}, rotate = 180] [color={rgb, 255:red, 0; green, 0; blue, 0 }  ][line width=0.75]    (10.93,-3.29) .. controls (6.95,-1.4) and (3.31,-0.3) .. (0,0) .. controls (3.31,0.3) and (6.95,1.4) .. (10.93,3.29)   ;
\draw    (290.22,103.66) -- (308.42,103.66) ;
\draw [shift={(310.42,103.66)}, rotate = 180] [color={rgb, 255:red, 0; green, 0; blue, 0 }  ][line width=0.75]    (10.93,-3.29) .. controls (6.95,-1.4) and (3.31,-0.3) .. (0,0) .. controls (3.31,0.3) and (6.95,1.4) .. (10.93,3.29)   ;
\draw    (378.93,103.66) -- (416.71,103.66) ;
\draw  [dash pattern={on 0.84pt off 2.51pt}] (186.56,64.67) .. controls (186.56,54.36) and (194.92,46) .. (205.23,46) -- (379.59,46) .. controls (389.9,46) and (398.26,54.36) .. (398.26,64.67) -- (398.26,120.69) .. controls (398.26,131) and (389.9,139.36) .. (379.59,139.36) -- (205.23,139.36) .. controls (194.92,139.36) and (186.56,131) .. (186.56,120.69) -- cycle ;
\draw    (161.97,72.09) -- (290.22,72.09) ;

\draw (320.86,92.17) node [anchor=north west][inner sep=0.75pt]  [font=\scriptsize] [align=left] {$\displaystyle \mathcal{D}_{\mathit{\tau_T}}( .) \ $};
\draw (339.62,93.29) node [anchor=north west][inner sep=0.75pt]   [align=left] {$ $};
\draw (208.72,93.36) node [anchor=north west][inner sep=0.75pt]  [font=\scriptsize] [align=left] {$\displaystyle \mathcal{A}^{*}( .\ ;\mathit{W_T}) \ $};
\draw (159,90) node [anchor=north west][inner sep=0.75pt]  [font=\scriptsize] [align=left] {$\displaystyle \mathbf{y}_{T-1}$};
\draw (418.82,94.73) node [anchor=north west][inner sep=0.75pt]  [font=\scriptsize] [align=left] {$\displaystyle X_{Temp}$};
\draw (166.57,57.66) node [anchor=north west][inner sep=0.75pt]  [font=\scriptsize] [align=left] {$\displaystyle \mathbf{b}$};

\end{tikzpicture}
\caption{Hidden Layer $T$}
\label{fig:hiddenlayerT}
\end{subfigure}
\caption{Deep neural network for the Learned Quantum State Tomography. The variables $\delta_0, \delta_t, \tau_t, \tau_T, W_t$ and  $W_T$ are learnable.}
\label{fig:DNN}
\end{figure*}

    \section{Learned Qunatum State Tomography (LQST) } \label{sec:LQST}

In this section, we design a deep neural network based on the SVT algorithm discussed in the previous section and present a training method to train the network for performing quantum state tomography.

\subsection{Network Architecture} \label{sec:architecture}

Recall that in each iteration of SVT \eqref{eqn:uzawaiterationsfinal}, two steps are performed. We first design a single hidden layer of our network which performs these two steps. Note that the SVT algorithm was mainly derived for estimating real matrices. To perform quantum state tomography, we use complex values for the matrices and vectors used in the algorithm. Precisely, the matrices $A_1$ through $A_m$ and the vector $\bm y_t$ are complex matrices and complex vectors $(A_1, \dots A_m \in \mathbb{C}^{d \times d}, \bm y_t \in \mathbb{C}^m)$. With this modification, the map $\mathcal{A}:\mathbb{C}^{d\times d} \to \mathbb{C}^m$ is defined as 
\begin{align}
    \mathcal{A}_i(X)=
    \tr[A_{i}X], \quad &\text{for }i=1,\dots m
    \label{eqn:qmap1}
\end{align} and the adjoint of the map $\mathcal{A}$ is given as
\begin{equation}
    \mathcal{A}^*(\bm y) = \sum_{i=1}^m y_i A_i^\dagger \label{eqn:adjoint1}
\end{equation}

We see that the steps \eqref{eqn:uzawaiterationsfinal} can be rewritten as
\begin{equation}
        \bm y_t = \bm y_{t-1} + \delta_t \left[\bm b - \mathcal{A}(\mathcal{D}_\tau(\mathcal{A}^*(\bm y_{t-1}))) \right] \label{eqn:svtunrolledeqn}
\end{equation} where $\bm y_t$ would represent the output of the $t^{th}$ hidden layer. It can be seen from this reformulation that performing \eqref{eqn:svtunrolledeqn} $T$ times is equivalent to running the SVT algorithm for $T$ iterations. We design the deep network wherein each hidden layer performs \eqref{eqn:svtunrolledeqn} \footnote{$\mathcal{D}_{\tau}(.)$ in \eqref{eqn:svtunrolledeqn} is the singular value thresholding (SVT) operator. Note that the recent work on RPCA \cite{RPCA} also designed a neural network which uses SVT in its layers and backpropagates through it}. To obtain a matrix as the output and to ensure that the output of the network meets the constraints required for a quantum state we add a few more layers at the end. 

To unroll a fixed number (say $T$) of SVT iterations, we build $T-1$ hidden layers each of which perform \eqref{eqn:svtunrolledeqn}. To ensure that the output of the network is a matrix, another layer is added which performs only the first step of \eqref{eqn:uzawaiterationsfinal}. Note that the output at this stage $(X_{temp})$ is a complex matrix and this may not meet the constraints required for the quantum state. For this purpose, we add an output layer with three sublayers $\mathcal{P}_{\mathcal{S}^d}(.), \mathcal{P}_{\mathcal{S}^{d^+}}(.)$ and $\mathcal{N}(.)$ as shown in Fig. \ref{fig:architechturelqst} to enforce the required constraints Hermitian, positive semidefiniteness and trace-one respectively. The notation $\mathcal{P}_{\mathcal{S}}(X)$ is used to denote the projection of the input $X$ onto the set $\mathcal{S}$. $\mathcal{S}^d$ and $\mathcal{S}^{d^+}$ denote the set of all $d$-dimensional complex Hermitian matrices and positive semidefinite matrices respectively. These layers were not used in the SVT algorithm as adopted for quantum state tomography in \cite{gross} because of the following reasons. Since SVT is a constrained optimization problem, the trace-one condition is incorporated as one of the linear constraints. Since the measurements matrices, $A_i$ used in \eqref{eqn:uzawaiterationsfinal} are all Hermitian, the SVT iterates $X^k$ are all Hermitian. Although the positive definiteness conditions are not explicitly imposed the output of the SVT algorithm is positive semidefinite with high probability. To show this, the empirical probability that the output of SVT is positive semidefinite is calculated in section \ref{sec:tuningsvt}. The output layer used in LQST is described subsequently.

The sublayer $\mathcal{P}_{\mathcal{S}^d}(X)$ \footnote{In our numerical simulations, it was noted that $X_{temp}$ is Hermitian and hence the sublayer $\mathcal{P}_{\mathcal{S}^d}(.)$ may not be necessary. However, this observation is solely based on empirical results. Since this layer is not computationally complex we have retained it as it does explicitly enforce the Hermitian property.} turns the input matrix $X$ Hermitian by computing $\frac{1}{2}(X + X^\dagger)$. $\mathcal{P}_{\mathcal{S}^{d^+}}(X)$ turns the input Hermitian matrix $X$ to positive semidefinite by performing eigen decomposition and setting all negative eigenvalues of $X$ to zero. $\mathcal{N}(X)$ enforces trace-one condition by normalizing the eigenvalues of $X$. Instead of normalizing the eigenvalues, the trace-one condition, in general, could be enforced by simply normalizing the input matrix as $\frac{1}{\tr[X]} X.$ We observed that when this way of normalization was incorporated, the gradients calculated during the backpropagation were diverging. Hence we added a small positive constant $\epsilon$ to the eigenvalues and performed normalization in the eigen domain for numerical stability. Note that we use the eigenvalues computed from the eigen decomposition performed to enforce positive definiteness rather than performing eigen decomposition again. The mathematical operations performed on $X_{temp}$ by the output layer are given as follows
\begin{align}
 X_{temp1} &\longleftarrow \frac{1}{2} \big( X_{temp} + X_{temp}^\dagger \big) \label{eqn:sublayer_hermitian} \\
        X_{temp2} &\longleftarrow X_{temp1} + \mu \, \text{diag}(1 \dots d) \label{eqn:sublayer_psdtraceone_start}\\
\text{Decompose: }X_{temp2} &= U \Lambda U^\dagger \nonumber \\
                  \hspace{1.1cm} &= U \text{diag}(\lambda_1 \dots \lambda_d) U^\dagger \\
                  \lambda_i &\longleftarrow \max(\lambda_i, 0) \hspace{3mm} i=1 \dots d \\     
                  \lambda_i &\longleftarrow \frac{ \lambda_i + \epsilon  }{\sum_{i=1}^{d} (\lambda_i + \epsilon)} \label{eqn:epsilonintroduction} \\
\text{Reconstruct: \hspace{1.9mm} } X_{out} &\longleftarrow U \text{diag}(\lambda_1 \dots \lambda_d) U^\dagger \label{eqn:sublayer_psdtraceone_end}
\end{align}
The sublayer $\mathcal{P}_{\mathcal{S}^d} (.)$ is described in \eqref{eqn:sublayer_hermitian}. The sublayers $\mathcal{P}_{\mathcal{S}^{d^+}}(.)$ and $\mathcal{N}(.)$ are collectively described in equations \eqref{eqn:sublayer_psdtraceone_start} through \eqref{eqn:sublayer_psdtraceone_end}. The importance of the small constant $\mu$ used in \eqref{eqn:sublayer_psdtraceone_start} will be discussed in greater detail in section \ref{sec:bellstate}.

\subsection{Network weights}

We denote by $W_t$ the measurements matrices $A_1$ through $A_m$ used in the maps $\mathcal{A}$ and $\mathcal{A}^*$ and make it learnable. Here $W_t$ is an $m \times d^2$ matrix whose $i^{th}$ row is formed from the entries of $A_i$ such that the measurement vector $\bm b$ given in \eqref{eqn:qmes} can be written as $W_t vec(X)$. By making this pair $\{\mathcal{A},\mathcal{A}^*\}$ learnable we try to leverage the power of deep learning to obtain an unrolled variant of SVT which performs quantum state tomography better than SVT (which has fixed known pair of $\{\mathcal{A}, \mathcal{A}^*\}$. We denote by $\delta_t$ the step size used in the $t^{th}$ layer and by $\tau_t$ the threshold used in the $t^{th}$ layer and we also make these learnable. With these learnable parameters, the complete network architecture is depicted in Fig. \ref{fig:architechturelqst} where Fig. \ref{fig:hiddenlayerTm1} and Fig. \ref{fig:hiddenlayerT} depict the hidden layers of our unrolled network.

\subsection{Training the network} \label{sec:training}

Consider a LQST network with $T$ layers as discussed in previous subsection. The input to this network is $\bm y_0 \in \mathbb{R}^m$ and the output of the network is a density matrix $\hat X \in \mathbb{C}^{d \times d}$. From \eqref{eqn:svtunrolledeqn} it can be seen that if $\bm y_{0}$ was a zero vector, $\bm y_1$ would then be $\delta \bm b$. Hence we feed the network with $\bm y_0 = \delta_0 \bm b$ to recover the corresponding matrix $X$. We denote by $\Theta$ the set of all learnable parameters in the network i.e., $\Theta = \{ W_1, \dots W_{T}, \delta_0, \delta_1, \dots \delta_{T-1}, \tau_1, \dots \tau_T \}$. With this notation, the output of the network for a given measurement vector (say $\bm b$) is written as
\begin{equation}
    \hat X = f_{\Theta}(\bm b)
\end{equation}

We denote the training dataset by $\{X^{(i)}, \bm b^{(i)}\}_{i=1}^M$ where $\bm b^{(i)} \in \mathbb{R}^m$ is the measurement vector of the matrix $X^{(i)} \in \mathbb{C}^{d \times d}$ obtained using the known measurement matrices $A_1$ through $A_m$ as described in \eqref{eqn:qmap}. For our numerical simulations, which we discuss in the next section, we generate the measurement matrices and $\{X^{(i)}\}$ synthetically. Then we obtain $\{\bm b^{(i)}\}$ using \eqref{eqn:qmap} as $\bm b^{(i)} = \mathcal{A}(X^{(i)} ; A_1, \dots A_m)$. We train our network to minimize the normalized mean squared error (NMSE) between the matrices $\{X^{(i)}\}$ in the training dataset and the estimated matrices $f_{\Theta}(\bm b^{(i)})$. The NMSE loss is given as
\begin{equation}
    \ell \{ \Theta;\{X^{(i)}, \bm b^{(i)}\} \} = \frac{1}{M d^2} \sum_{i=1}^M \norm{ X^{(i)} - f_{\Theta}(\bm b^{(i)}) }_F^2 \label{eqn:mse}
\end{equation}

\begin{table*}[ht!]
\centering
\resizebox{13cm}{!}{
\begin{tabular}{|c|c|c|c|c|c|c|c|c|}
\hline
\multirow{2}{*}{\textbf{$\tau$}} & \multicolumn{2}{c|}{\textbf{$\delta = 0.01$}} & \multicolumn{2}{c|}{\textbf{$\delta = 0.1$}} & \multicolumn{2}{c|}{\textbf{$\delta = 0.5$}} & \multicolumn{2}{c|}{\textbf{$\delta = 2.982$}} \\ \cline{2-9} 
                                 & \textbf{Iterations}    & \textbf{Fidelity}    & \textbf{Iterations}    & \textbf{Fidelity}   & \textbf{Iterations}    & \textbf{Fidelity}   & \textbf{Iterations}     & \textbf{Fidelity}    \\ \hline
\textbf{2}                       & 14783                  & 0.8773               & \textbf{1632}                   & 0.8797              & -1                     & -1                  & -1                      & -1                   \\ \hline
\textbf{4}                       & 17939                  & 0.8738               & 3195                   & 0.8746              & 20000                  & 0.6695              & -1                      & -1                   \\ \hline
\textbf{6}                       & 18752                  & 0.8786               & 4785                   & 0.8799              & 20000                  & 0.6596              & -1                      & -1                   \\ \hline
\textbf{8}                       & 19419                  & 0.8927               & 4694                   & 0.8653              & 20000                  & 0.6476              & -1                      & -1                   \\ \hline
\textbf{10}                      & 19753                  & 0.9094               & 6314                   & 0.8751              & 20000                  & 0.6526              & -1                      & -1                   \\ \hline
\textbf{80}                      & 20000                  & 0.9349               & 19054                  & 0.9166              & 20000                  & 0.6625              & -1                      & -1                   \\ \hline
\end{tabular}
}
\caption{SVT for estimating quantum state of rank-3 is simulated with different tuning parameters. }
\label{tab:tuningsvt}
\end{table*}

\begin{table*}[ht!]
\centering
\resizebox{18cm}{!}{
\begin{tabular}{|c|ccc|cccccc|}
\hline
\multirow{3}{*}{\textbf{Rank}} &
  \multicolumn{3}{c|}{\multirow{2}{*}{\textbf{\begin{tabular}[c]{@{}c@{}}SVT\\ $\tau = 2, \delta = 0.1$\end{tabular}}}} &
  \multicolumn{6}{c|}{\textbf{LQST}} \\ \cline{5-10} 
 &
  \multicolumn{3}{c|}{} &
  \multicolumn{2}{c|}{\textbf{$T=2$}} &
  \multicolumn{2}{c|}{\textbf{$T=3$}} &
  \multicolumn{2}{c|}{\textbf{$T=4$}} \\ \cline{2-10} 
 &
  \multicolumn{1}{c|}{\textbf{Iterations}} &
  \multicolumn{1}{c|}{\textbf{Fidelity}} &
  \textbf{Trace dist.} &
  \multicolumn{1}{c|}{\textbf{\begin{tabular}[c]{@{}c@{}}Fidelity\\ Std. Dev.\end{tabular}}} &
  \multicolumn{1}{c|}{\textbf{\begin{tabular}[c]{@{}c@{}}Trace dist.\\ Std. Dev.\end{tabular}}} &
  \multicolumn{1}{c|}{\textbf{\begin{tabular}[c]{@{}c@{}}Fidelity\\ Std. Dev.\end{tabular}}} &
  \multicolumn{1}{c|}{\textbf{\begin{tabular}[c]{@{}c@{}}Trace dist.\\ Std. Dev.\end{tabular}}} &
  \multicolumn{1}{c|}{\textbf{\begin{tabular}[c]{@{}c@{}}Fidelity\\ Std. Dev.\end{tabular}}} &
  \textbf{\begin{tabular}[c]{@{}c@{}}Trace dist.\\ Std. Dev.\end{tabular}} \\ \hline
\multirow{2}{*}{\textbf{3}} &
  \multicolumn{1}{c|}{\multirow{2}{*}{1632}} &
  \multicolumn{1}{c|}{\multirow{2}{*}{0.8751}} &
  \multirow{2}{*}{0.3078} &
  \multicolumn{1}{c|}{0.894} &
  \multicolumn{1}{c|}{0.3488} &
  \multicolumn{1}{c|}{0.9171} &
  \multicolumn{1}{c|}{0.3053} &
  \multicolumn{1}{c|}{\textbf{0.923}} &
  \textbf{0.2884} \\ \cline{5-10} 
 &
  \multicolumn{1}{c|}{} &
  \multicolumn{1}{c|}{} &
   &
  \multicolumn{1}{c|}{0.029} &
  \multicolumn{1}{c|}{0.0429} &
  \multicolumn{1}{c|}{0.024} &
  \multicolumn{1}{c|}{0.0419} &
  \multicolumn{1}{c|}{0.0249} &
  0.0436 \\ \hline
\multirow{2}{*}{\textbf{4}} &
  \multicolumn{1}{c|}{\multirow{2}{*}{438}} &
  \multicolumn{1}{c|}{\multirow{2}{*}{0.8115}} &
  \multirow{2}{*}{0.4213} &
  \multicolumn{1}{c|}{0.8688} &
  \multicolumn{1}{c|}{0.3856} &
  \multicolumn{1}{c|}{0.8893} &
  \multicolumn{1}{c|}{0.3534} &
  \multicolumn{1}{c|}{\textbf{0.8957}} &
  \textbf{0.3394} \\ \cline{5-10} 
 &
  \multicolumn{1}{c|}{} &
  \multicolumn{1}{c|}{} &
   &
  \multicolumn{1}{c|}{0.0281} &
  \multicolumn{1}{c|}{0.0393} &
  \multicolumn{1}{c|}{0.0255} &
  \multicolumn{1}{c|}{0.0406} &
  \multicolumn{1}{c|}{0.0253} &
  0.0419 \\ \hline
\multirow{2}{*}{\textbf{5}} &
  \multicolumn{1}{c|}{\multirow{2}{*}{226}} &
  \multicolumn{1}{c|}{\multirow{2}{*}{0.7973}} &
  \multirow{2}{*}{0.4457} &
  \multicolumn{1}{c|}{0.856} &
  \multicolumn{1}{c|}{0.4032} &
  \multicolumn{1}{c|}{0.8724} &
  \multicolumn{1}{c|}{0.3796} &
  \multicolumn{1}{c|}{\textbf{0.8775}} &
  \textbf{0.3701} \\ \cline{5-10} 
 &
  \multicolumn{1}{c|}{} &
  \multicolumn{1}{c|}{} &
   &
  \multicolumn{1}{c|}{0.0254} &
  \multicolumn{1}{c|}{0.0347} &
  \multicolumn{1}{c|}{0.0246} &
  \multicolumn{1}{c|}{0.0367} &
  \multicolumn{1}{c|}{0.0248} &
  0.0381 \\ \hline
\end{tabular}
}
\caption{Comparing SVT and LQST}
\label{tab:comparingsvtlqst}
\end{table*}

We initialize the network's trainable weights $\{W_t\}$ with the measurement matrices $A_1$ through $A_m$. Each of the stepsizes $\{\delta_0, \delta_1, \dots \delta_T\}$ and the thresholds $\{\tau_1, \dots \tau_T\}$ is initialized to $0.01$. The two constants $\mu$ and $\epsilon$ used in the output layer to numerically stabilize the gradients are set to 0 and $10^{-8}$ respectively. We use gradient-based optimizer ADAM \cite{adam} to minimize the loss \eqref{eqn:mse} and train the network.

    \section{Numerical simulation} \label{sec:simulation}

We design and train the proposed Learned Quantum State Tomography network to estimate the quantum state of $4$-qubit system described by a $d \times d$ matrix of rank - $r$ by observing Pauli measurements as described in section \ref{sec:problem}. Here, $d=2^4=16$. Only 40 percent of the total measurements required for the tomographically complete set of measurements are observed i.e., the number of measurements $m = 103 \simeq 0.4 \times (d^2-1)$. The true density matrix $\rho^* \in \mathbb{C}^{16 \times 16}$ of rank $r$ is generated as $\rho^* = G G^\dagger / \tr[G G^\dagger] $ where the real and imaginary entries of $G \in \mathbb{C}^{16 \times r}$ are randomly generated from standard normal distribution, $\mathcal{N}(0,1)$. This generation ensures that $\rho^*$ is unit trace and positive semidefinite. To obtain the measurements of the state, we first choose $m$ observables uniformly at random from the first $d^2$ observables in $\mathcal{O}$ as described in section \ref{sec:problem}. Then, the measurements of the true state are obtained using \eqref{eqn:qmap} and these measurements are used to estimate the true state. 

We first tune the SVT algorithm by simulating it for different values of the tunable parameters $\tau$ and $\delta$ for faster convergence of the algorithm. Then we compare the tuned SVT against our proposed LQST.

\subsection{Tuning SVT} \label{sec:tuningsvt}

To find the best tunable parameters $(\tau \text{ and } \delta)$, SVT is simulated to estimate the state of $4$-qubit system of different ranks with different tunable parameters. Note that the recommended value for $\tau$ and $\delta$ suggested by the authors of SVT \cite{svt} are $5*d=80$ and $1.2 \times d^2/m = 2.982$. To tune SVT, the set of values used for $\tau$ is 2,4,6,8,10 and 80 and the set of values used for $\delta$ is 0.01, 0.1, 0.5 and 2.982. The number of iterations and fidelity in estimating rank-3 state by simulating SVT until convergence with a maximum allowable iterations of 20000 is tabulated in Table. \ref{tab:tuningsvt}. We use the convergence criterion suggested by the authors of SVT given as $\frac{ \norm{ \mathcal{A}(X_k) - b}_F }{\norm{b}_F} < 10^{-4}.$ The entries $-1$ in Table. \ref{tab:tuningsvt} denote that the algorithm diverges for the corresponding values of $\tau$ and $\delta$. It can be seen from Table. \ref{tab:tuningsvt} that the SVT converges faster when $\tau = 2$ and $\delta = 0.1$ are used. We used these parameters to compare SVT against our proposed LQST.

As discussed in section \ref{sec:architecture}, the output of SVT is not necessarily positive semidefinite. The empirical probability that the output of SVT is positive semidefinite (PSD) is calculated by performing the quantum estimation of the rank-3 4-qubit system 1000 times and counting the number of times the output was PSD. The probability is computed for different values of threshold $(\tau)$ and stepsize $(\delta)$ used in SVT and are tabulated in Table \ref{tab:svt_ispd}. The entries -1 in Table \ref{tab:svt_ispd} denote that SVT diverges for the corresponding values of $\tau$ and $\delta$. It is observed that the output of SVT is positive semidefinite with high probability.

\begin{table}[H]
\centering
\begin{tabular}{|c|cccc|}
\hline
\multirow{2}{*}{\textbf{$\tau$}} & \multicolumn{4}{c|}{\textbf{Empirical probabilities}}                                     \\ \cline{2-5} 
 & \multicolumn{1}{c|}{\textbf{$\delta=0.01$}} & \multicolumn{1}{c|}{\textbf{$\delta=0.1$}} & \multicolumn{1}{c|}{\textbf{$\delta=0.5$}} & \textbf{$\delta=2.982$} \\ \hline
\textbf{2}                       & \multicolumn{1}{c|}{0.987} & \multicolumn{1}{c|}{0.965} & \multicolumn{1}{c|}{-1}    & -1 \\ \hline
\textbf{4}                       & \multicolumn{1}{c|}{1}     & \multicolumn{1}{c|}{1}     & \multicolumn{1}{c|}{0.732} & -1 \\ \hline
\textbf{6}                       & \multicolumn{1}{c|}{1}     & \multicolumn{1}{c|}{1}     & \multicolumn{1}{c|}{0.912} & -1 \\ \hline
\textbf{8}                       & \multicolumn{1}{c|}{1}     & \multicolumn{1}{c|}{1}     & \multicolumn{1}{c|}{0.96}  & -1 \\ \hline
\textbf{10}                      & \multicolumn{1}{c|}{1}     & \multicolumn{1}{c|}{1}     & \multicolumn{1}{c|}{0.988} & -1 \\ \hline
\textbf{80}                      & \multicolumn{1}{c|}{1}     & \multicolumn{1}{c|}{1}     & \multicolumn{1}{c|}{1}     & -1 \\ \hline
\end{tabular}
\caption{Probability that the output of SVT is positive semidefinite.}
\label{tab:svt_ispd}
\end{table}

\begin{figure*}[ht!]
\hspace{9mm}

    \begin{subfigure}[H]{.3\textwidth}
        \centering
        \includegraphics[width=\textwidth]{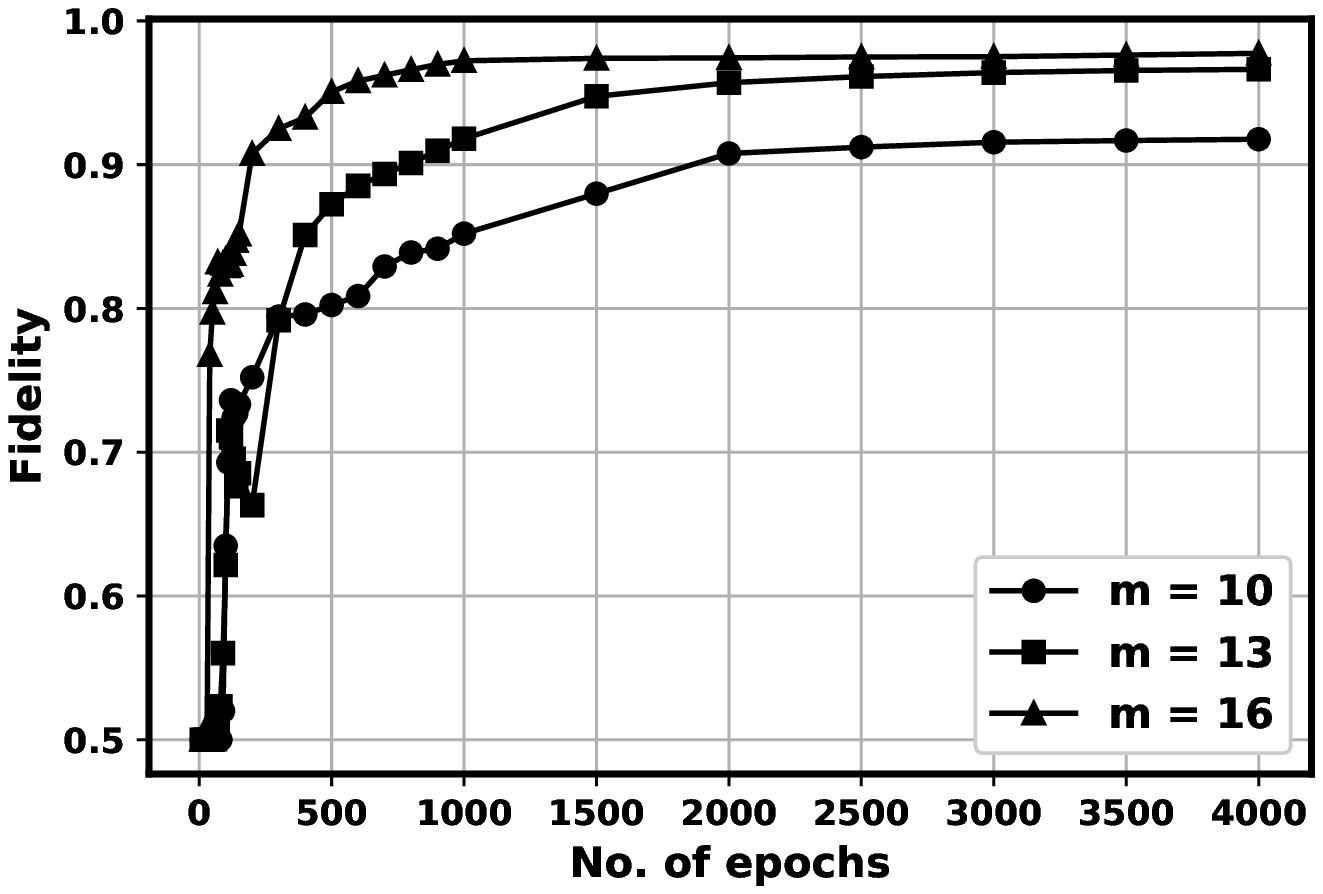}
        \caption{Fidelity}
        \label{fig:bell_trace_diffm_T3_navg1000_noclr_fd}
    \end{subfigure}
\hfill
    \begin{subfigure}[H]{.3\textwidth}
        \centering
        \includegraphics[width=\textwidth]{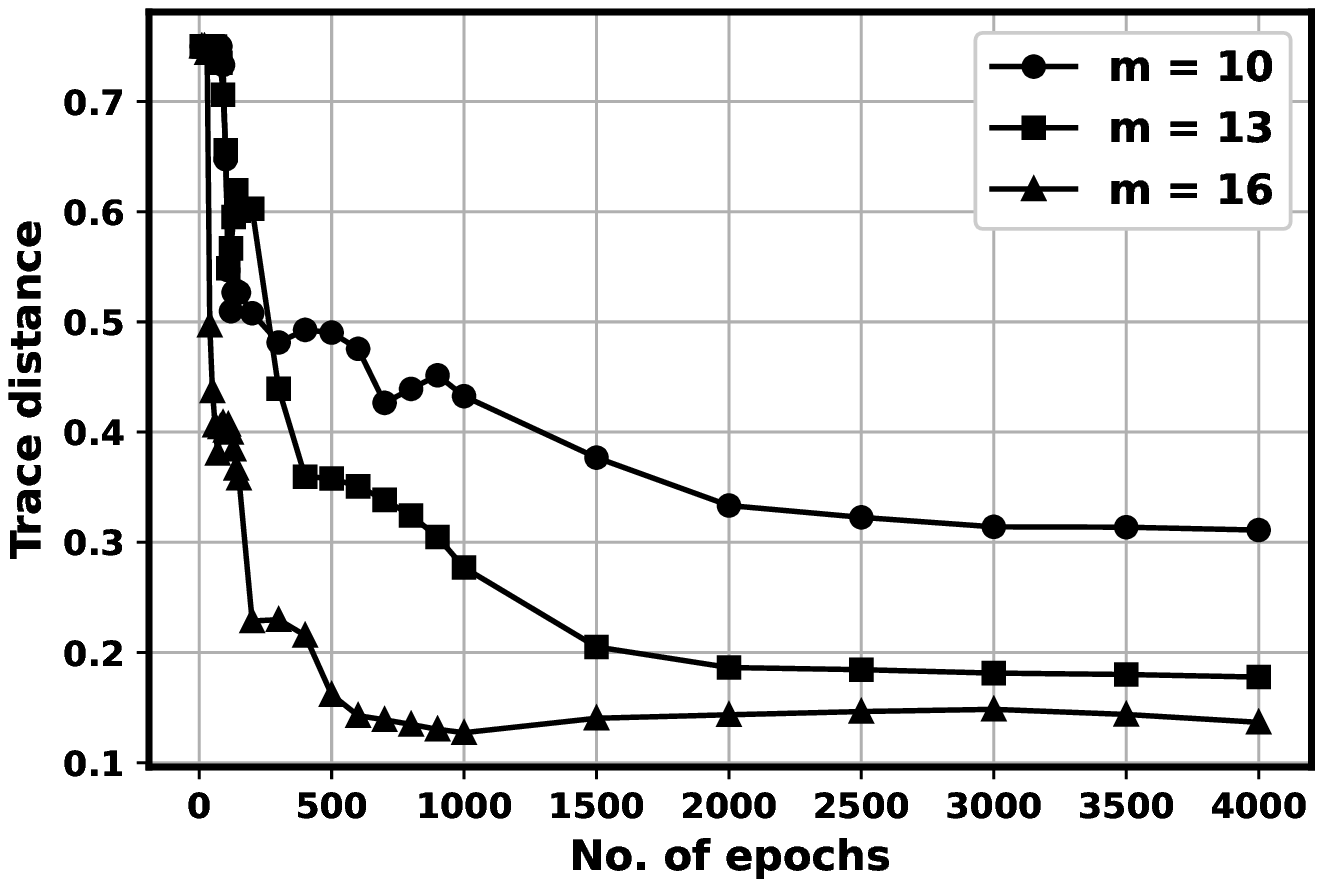}
        \caption{Trace Distance}
        \label{fig:bell_trace_diffm_T3_navg1000_noclr_td}
    \end{subfigure}
\hfill
    \begin{subfigure}[H]{.3\textwidth}
        \centering
        \includegraphics[width=\textwidth]{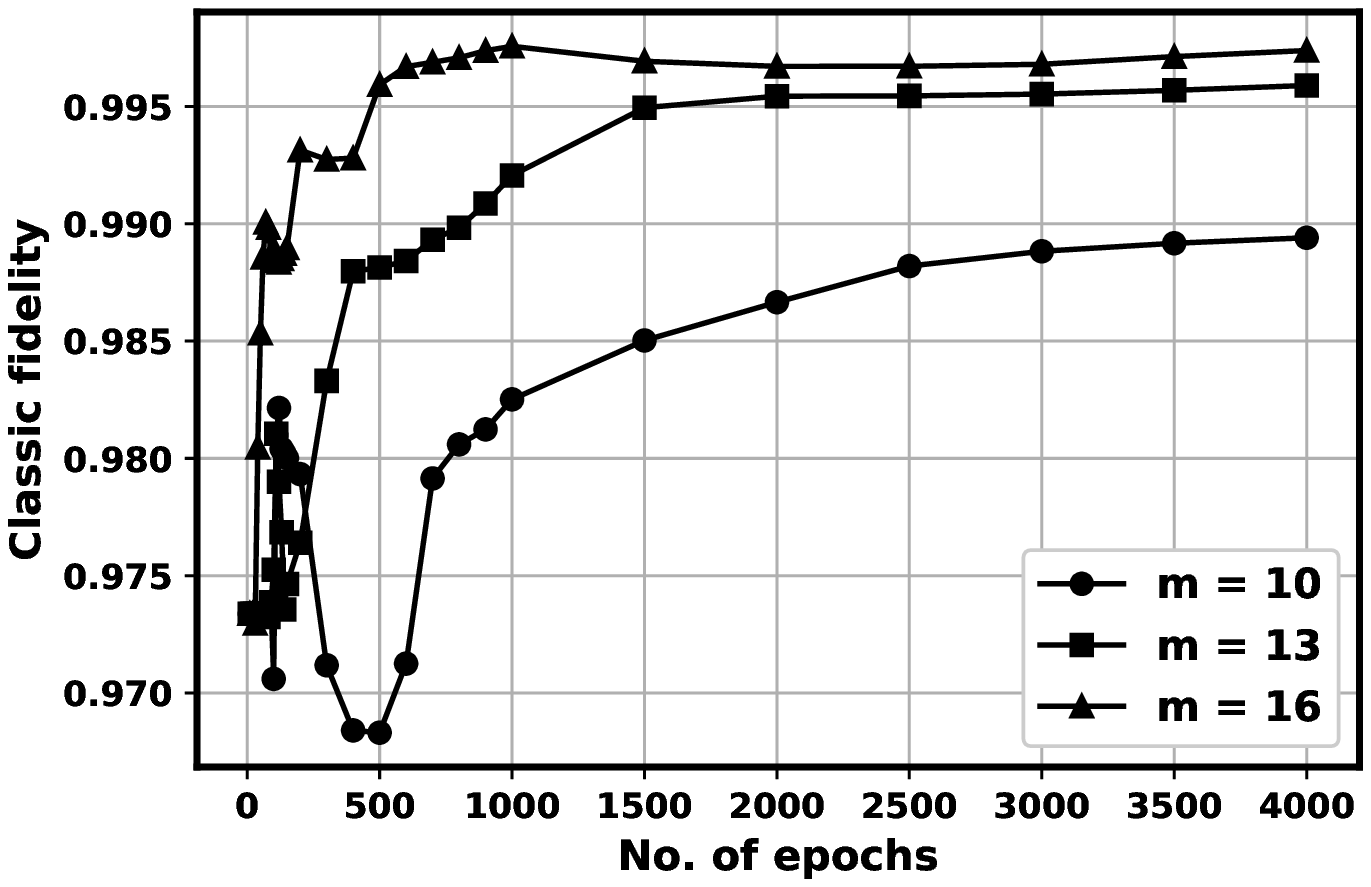}
        \caption{Classic fidelity}
        \label{fig:bell_trace_diffm_T3_navg1000_noclr_cf}
    \end{subfigure}

\caption{Comparing the performance of LQST in estimatimg the Bell state by varying the number of POVM probabilities $m$ used in estimation.}
\label{fig:bell_trace_diffm_T3_navg1000_noclr}
\end{figure*}

\begin{figure*}[ht!]
\hspace{9mm}

    \begin{subfigure}[H]{.3\textwidth}
        \centering
        \includegraphics[width=\textwidth]{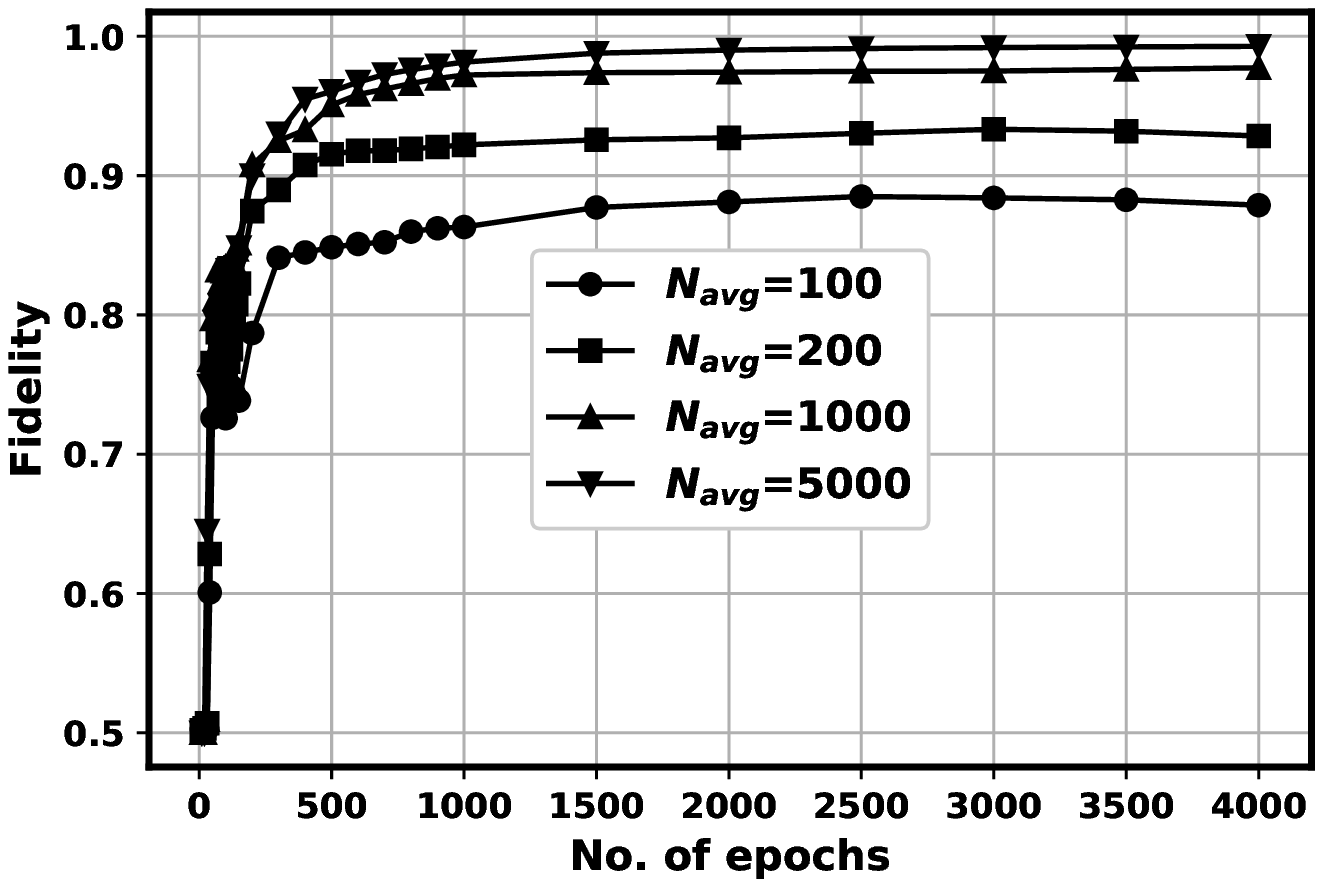}
        \caption{Fidelity}
        \label{fig:bell_trace_diffnavg_m16_T3_noclr_fd}
    \end{subfigure}
\hfill
     \begin{subfigure}[H]{.3\textwidth}
        \centering
        \includegraphics[width=\textwidth]{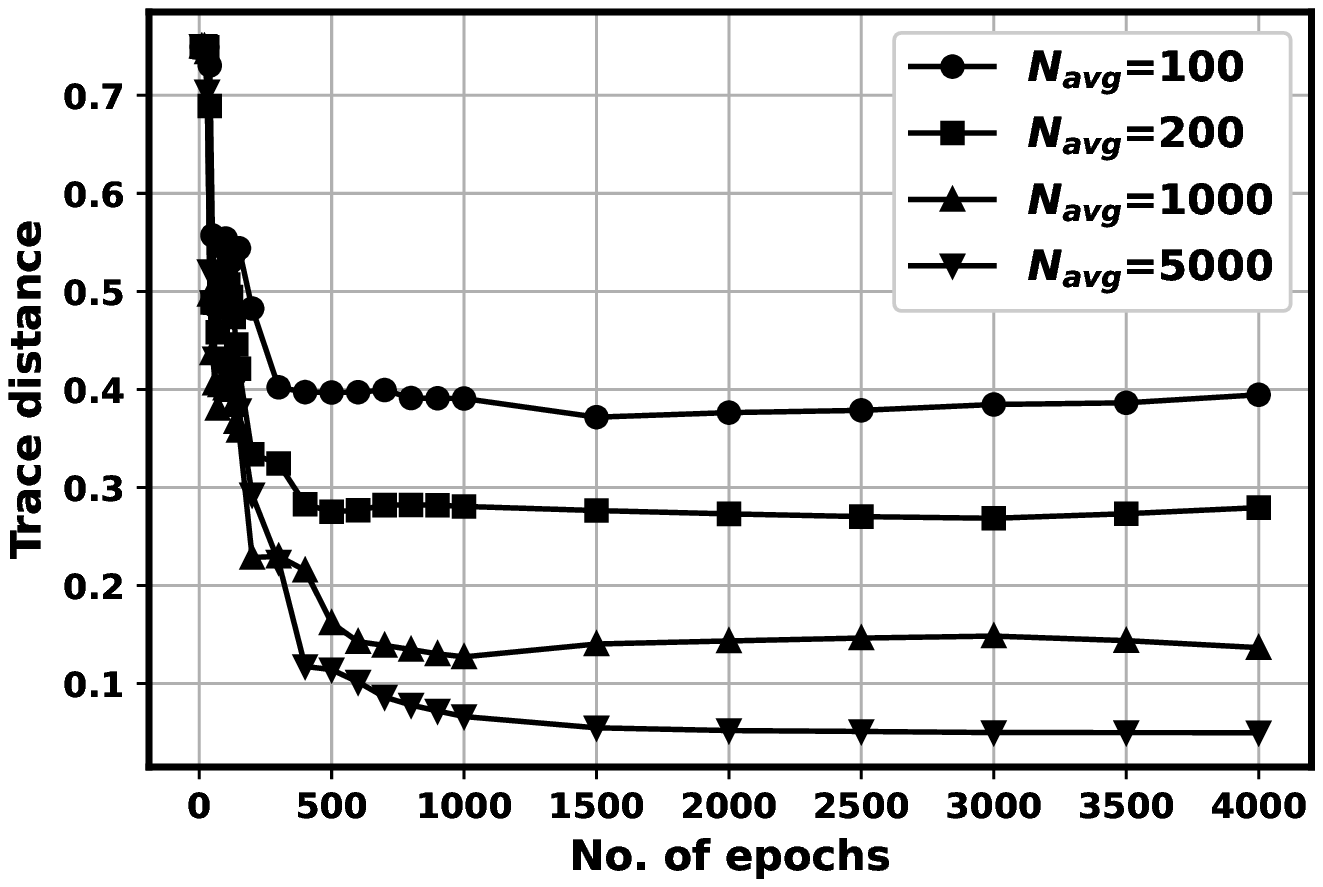}
        \caption{Trace distance}
        \label{fig:bell_trace_diffnavg_m16_T3_noclr_td}
    \end{subfigure}
\hfill
     \begin{subfigure}[H]{.3\textwidth}
        \centering
        \includegraphics[width=\textwidth]{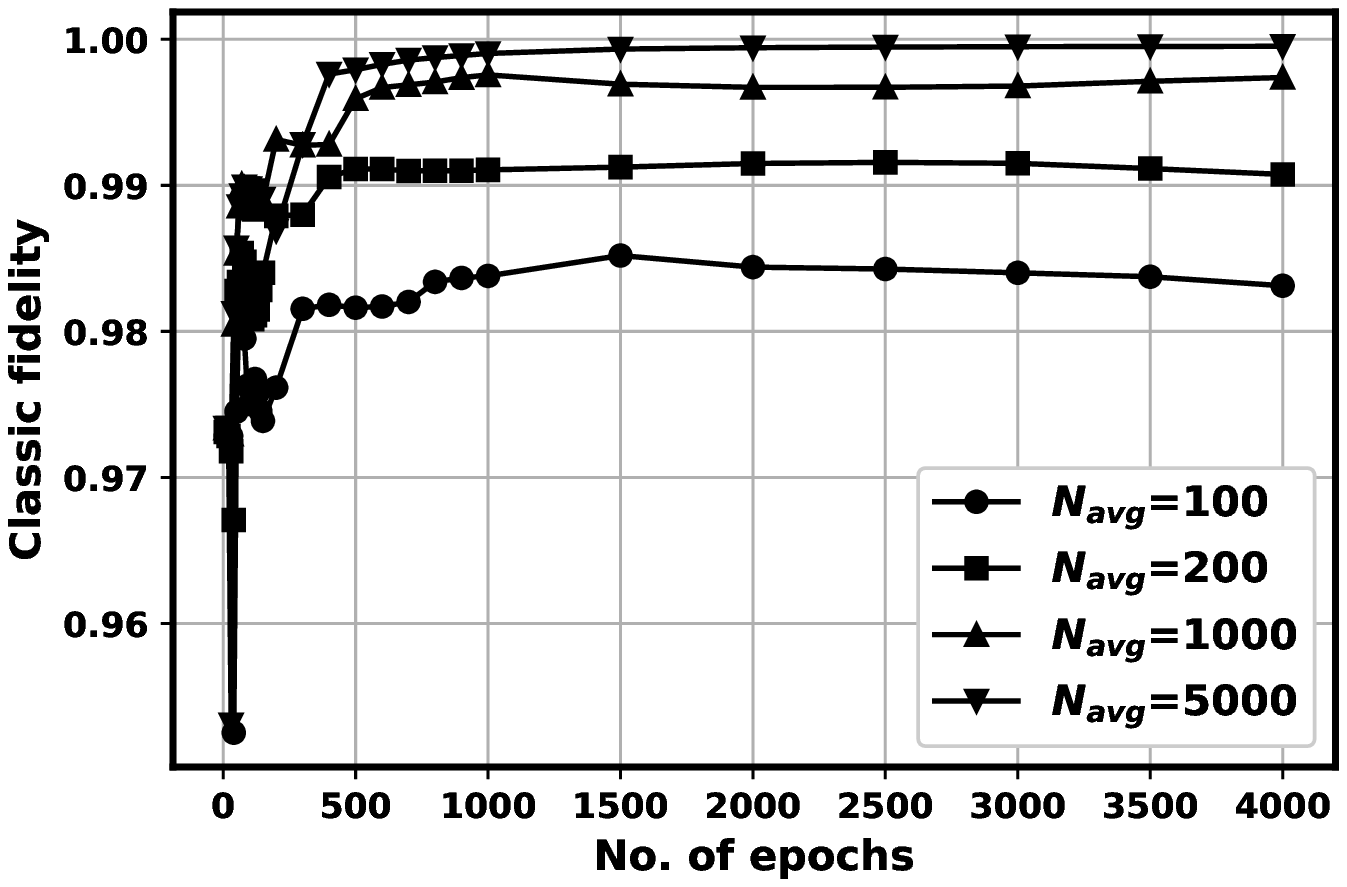}
        \caption{Classic fidelity}
        \label{fig:bell_trace_diffnavg_m16_T3_noclr_cf}
    \end{subfigure}

    \begin{subfigure}[H]{.3\textwidth}
        \centering
        \includegraphics[width=\textwidth]{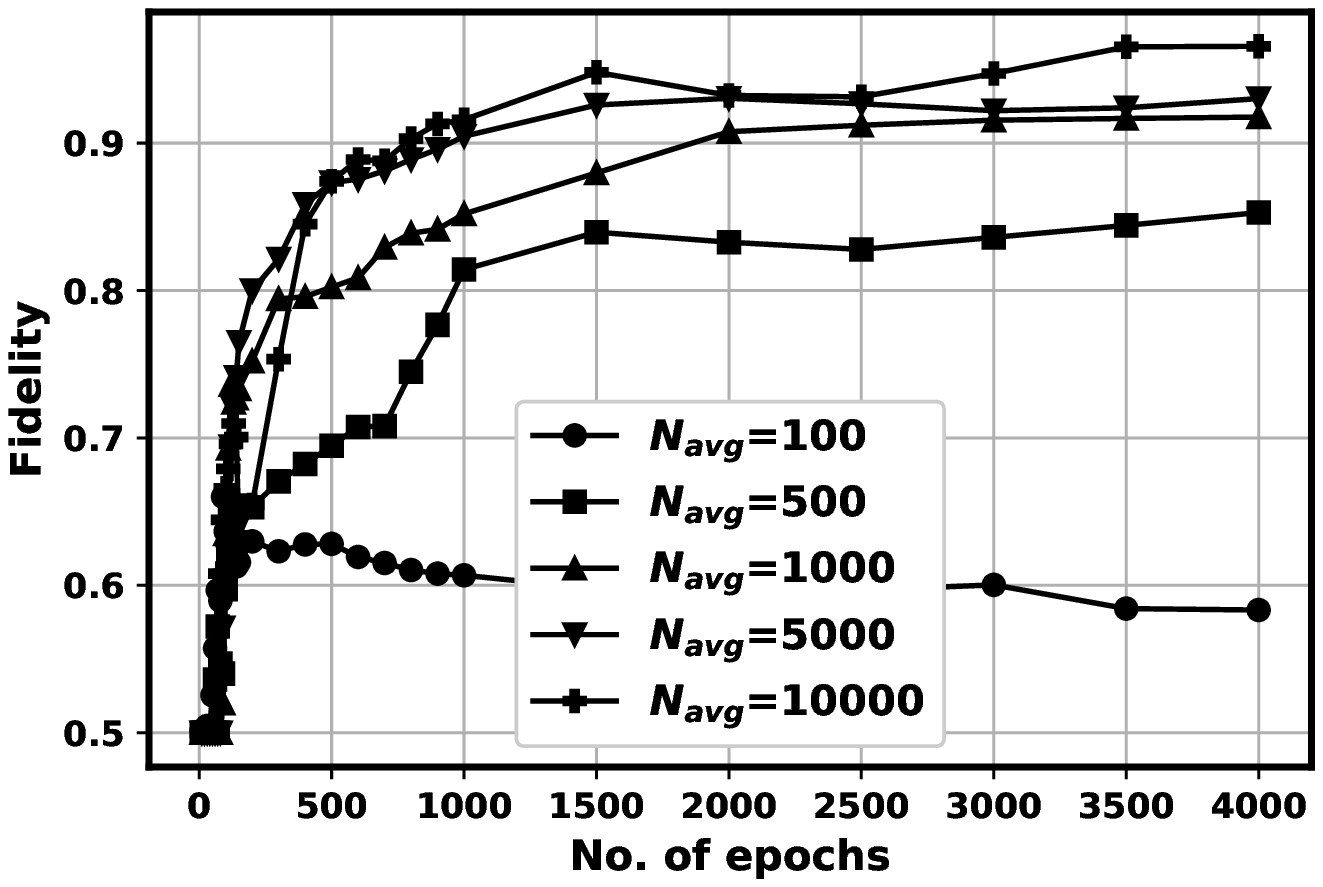}
        \caption{Fidelity}
        \label{fig:bell_trace_diffnavg_m10_T3_noclr_fd}
    \end{subfigure}
\hfill
     \begin{subfigure}[H]{.3\textwidth}
        \centering
        \includegraphics[width=\textwidth]{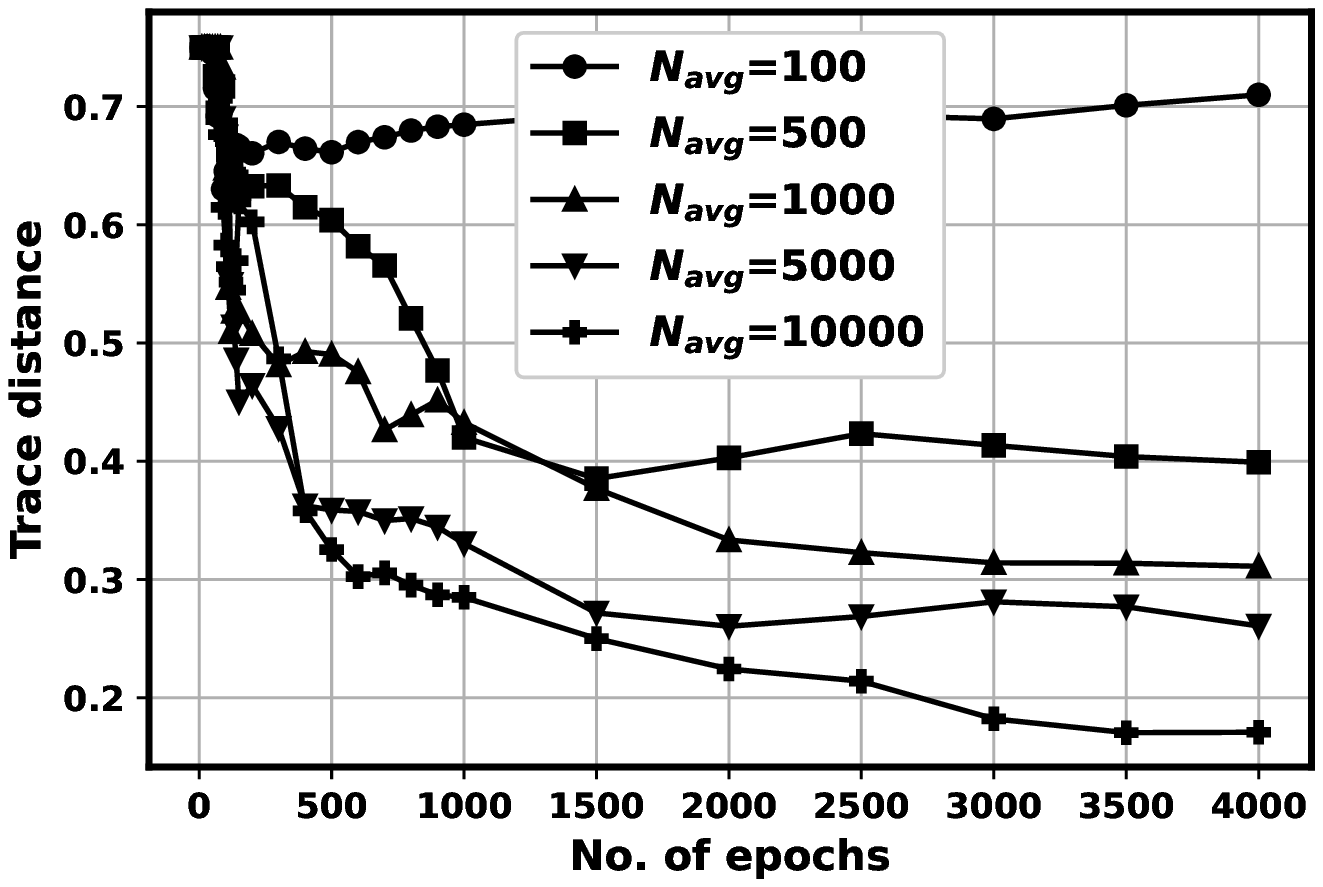}
        \caption{Trace distance}
        \label{fig:bell_trace_diffnavg_m10_T3_noclr_td}
    \end{subfigure}
\hfill
     \begin{subfigure}[H]{.3\textwidth}
        \centering
        \includegraphics[width=\textwidth]{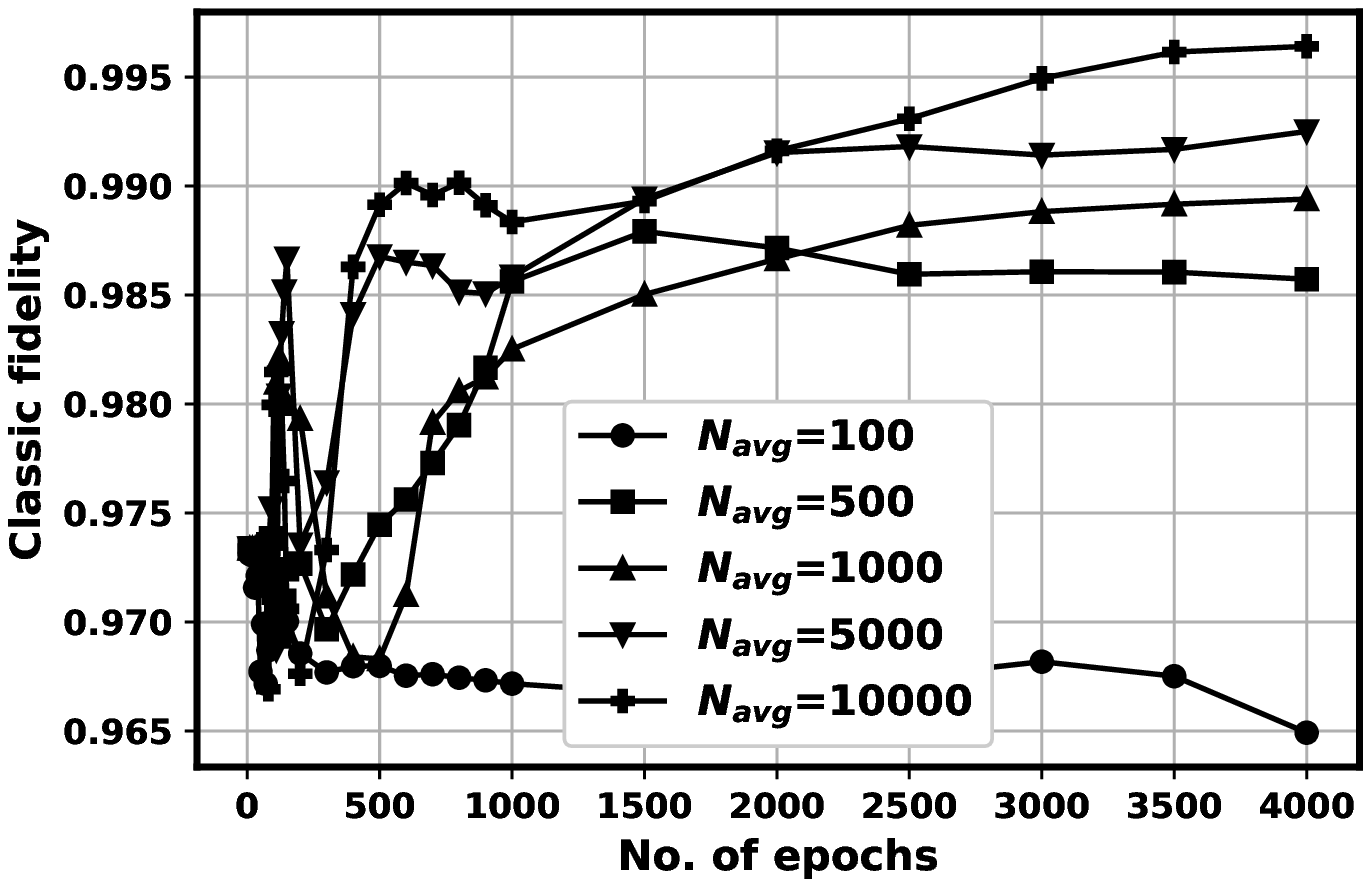}
        \caption{Classic fidelity}
        \label{fig:bell_trace_diffnavg_m10_T3_noclr_cf}
    \end{subfigure}

\caption{Comparing the estimation of Bell state by varying the number of individual measurements used in the calculation of POVM probabilities. (\subref{fig:bell_trace_diffnavg_m16_T3_noclr_fd}),(\subref{fig:bell_trace_diffnavg_m16_T3_noclr_td}),(\subref{fig:bell_trace_diffnavg_m16_T3_noclr_cf}): Performance of LQST when an innformationally complete set of measurements is observed i.e., $m=d^2=16$.  (\subref{fig:bell_trace_diffnavg_m10_T3_noclr_fd}),(\subref{fig:bell_trace_diffnavg_m10_T3_noclr_td}),(\subref{fig:bell_trace_diffnavg_m10_T3_noclr_cf}): Performance of LQST when an innformationally incomplete set of measurements is observed i.e., $m=10$. }
\label{fig:bell_trace_diffnavg_m10_T3_noclr}
\end{figure*}

\subsection{Comparing SVT and LQST} \label{sec:comparingsvtandlqst}

We design and train the proposed LQST network to estimate 4-qubit state of rank $r$ from its $m=103$ linear measurements. To train the LQST network, 70000 ground-truth data $\{X^{(i)} \in \mathbb{C}^{16 \times 16}, \bm b^{(i)} \in \mathbb{R}^{103}\}_{i=1}^{70000}$ were generated as described in section \ref{sec:training} $\{X^{(i)}\}$ are the quantum states of rank-$r$ and $\{\bm b^{(i)}\}$ are the corresponding measurement vectors.  We use PyTorch to design and train our LQST network. PyTorch automatically computes the gradients of the loss function with respect to the network parameters using computational graphs and Autograd functionality. We use a stochastic gradient descent-based ADAM optimizer with a learning rate of $10^{-4}$ to train the network. We perform mini-batch training with a mini-batch size of 1000.  Of the 70,000 ground truth data, we use 50,000 to train the network, and 10,000 to validate the network. We validate the network every time the network parameter gets updated and we stop the training process when this validation loss doesn't decrease over the course of training. Once the training is over, we freeze the network parameters and test the network with the remaining 10000 ground truth data, and the resulted fidelity and trace distance (TD) values are reported along with their standard deviation. These values are compared against the Fidelity and Trace Distance in estimation by SVT averaged over 10000 instances and are tabulated in Table. \ref{tab:comparingsvtlqst}. The best fidelity and trace distance values for a given rank are in bold font.

It can be seen from Table. \ref{tab:comparingsvtlqst} that our proposed LQST with only a very few layers recovers the quantum state with better fidelity than SVT with many hundreds of iterations. In order to perform a quantum estimation with a fixed fidelity, our LQST would be the best choice as it recovers the state with much lesser computational complexity. For instance, to estimate a rank-3 $4$-qubit system with fidelity of 0.91, it can be seen from Table. \ref{tab:comparingsvtlqst} that LQST requires only 3 layers in its architecture. But SVT converges only after 1632 iterations yet fails to recover the state with such high fidelity.  Although it can be seen from Table. \ref{tab:tuningsvt} that SVT recovers the state with 0.91 fidelity when $\tau = 80$ and $\delta = 0.1$ are used, it takes 19054 iterations to converge. 

The output of SVT is not necessarily positive semidefinite whereas the output of LQST is always a valid quantum state since we enforce the required conditions for a quantum state on the output matrix by the addition of the output layer as discussed in section \ref{sec:architecture}. To see how the addition of $\epsilon=10^{-8}$ in the output layer affects the rank of the estimated matrix, the rank of the matrices estimated by SVT and LQST are compared in Table \ref{tab:rank_comparision}. The rank of the output matrix $X$ is calculated as the number of singular values of $X$ that are greater than a threshold of $10^{-7}$. It is observed that the rank estimated by LQST is very close to the rank of the underlying matrix denoted as $r$ and is closer than that of SVT.

\begin{table}[H]
\centering
\begin{tabular}{|c|c|c|}
\hline
\textbf{rank ($r$)} & \textbf{LQST} & \textbf{SVT} \\ \hline
3 & 4.2550 & 7.112 \\ \hline
4 & 5.3020 & 8.812 \\ \hline
5 & 6.2450 & 10.169 \\ \hline
\end{tabular}
\caption{Comparing the rank of the output matrix estimated by LQST with that of SVT.}
\label{tab:rank_comparision}
\end{table}

We see that to perform quantum state tomography with a fixed fidelity our LQST will be a much better choice as its computational complexity is very much lesser than SVT. However, note that the claim that our method is computationally more efficient than SVT does not take into account the training procedure and the network has to be retrained for a different set of observables. However, once the network is trained for a given set of measurements, it uses only very few layers to estimate the state. Further, our network is more efficient than the algorithms which use the Bayesian approach such as \cite{bayesianinference}. In \cite{bayesianinference}, the authors have parameterized the density matrix of size $d \times d$ into $d^2-1$ parameters, and these parameters are estimated using Bayesian inference. The density matrix is estimated by mapping the expectation of the posterior distribution of the $d^2 - 1$ parameters to the set of density matrices. To calculate the expectation of the posterior distribution of these parameters, the authors have used the Metropolis-Hastings algorithm (a Markov Chain Monte Carlo - (MCMC) algorithm ) which requires a few thousand iterations. On the contrary, our trained network uses only a few layers to estimate the density matrix. If quantum state tomography were to be performed using low-power devices LQST would be a much better choice as it takes only a very few layers to produce the same result as SVT would produce in many thousands of iterations.

\subsection{Estimation of Bell state} \label{sec:bellstate}

In the previous subsection, we demonstrated the estimation of random quantum states using our method from noiseless measurements which were the expectation values of the  Pauli observables as given in \eqref{eqn:qmes}. In this section, following \cite{rqst_gm} and \cite{nnqst2qubit}, we train our network to estimate quantum Bell state $\rho_{Bell} \in \mathbf{C}^{d \times d}$ (given in Eqn. \eqref{eqn:rhobell}) which is a 2-qubit quantum state with $d=4$ from an informationally complete POVM (Positive Operator Valued Measure) \cite{qcqi} measurements, i.e., $m=d^2=16$. We also demonstrate the estimation of Bell state from informationally incomplete set of measurements. We use noisy POVM measurements by performing finite number of individual quantum measurements rather than using noiseless expectation values of the observables as used in the previous subsection. We use modified Pauli POVM (Pauli-4 POVM) as used by the authors of \cite{nnqst2qubit} and \cite{rqst_gm}. Four positive semidefinite operators $\{ M_a^{(1)} \in \mathbb{C}^{2 \times 2} \}_{a \in \{1,2,3,4\}}$ which define the Pauli-4 POVM on 1-qubit sytem are explicitly given in \cite{nnqst2qubit}. Pauli-4 POVM on 2-qubit system is defined by the 16 product operators $\{ M_a^{(2)} = M_{a_1}^{(1)} \otimes M_{a_2}^{(1)} \}_{a_1, a_2 \in \{1,2,3,4\} , a \in \{1,\dots 16\} }$ which correspond to 16-outcome POVM measurement of the 2-qubit system. The exact probability that the outcome of the measurement being $a$ is given by

\begin{align}
    P_{Bell}(a) = Tr[\rho_{Bell} M_a^{(2)}]  \label{eqn:povmprob}
\end{align}

We call these probabilities as POVM-probabilities.  To get the noisy value of these probabilities, we perform POVM measurement on Bell state for $N_{avg}$ times. The noisy value of $P_{Bell}(a)$ for a given $a$ is then calculated by $N_a / N_{avg}$ where $N_a$ is the number of times the outcome is $a$.

\begin{align}
    \rho_{Bell} = \begin{bmatrix} 0.5 & 0 & 0 & 0.5 \\ 0 & 0 & 0 & 0 \\  0 & 0 & 0 & 0 \\ 0.5 & 0 & 0 & 0.5 \end{bmatrix} \label{eqn:rhobell}
\end{align}

The training data set is created by randomly generating $M(=600)$ rank-1 density matrices $\{ X^{(i)} \}_{i=1}^M \in \mathbb{C}^{4 \times 4}$ as described in section \ref{sec:training} and the corresponding noisy values of POVM-probabilities $\{ P_{X^{(i)}}(\bm{a}) \}_{i=1}^M$  are obtained. We split these 600 training samples into 500 training data and 100 validation data to train and validate the network respectively. 3-layered LQST is trained on this dataset with the mini-batch size of 50 and a learning rate of $10^{-4}$. It was observed while estimating the quantum Bell state that the gradients computed were not numerically stable when any two eigenvalues of
the matrix to be decomposed are the same. To overcome this, a small positive constant $i \mu$ was added to the $i^{th}$ eigenvalue of $X$ 
before the eigen decomposition to numerically stabilize the gradients as given in \eqref{eqn:sublayer_psdtraceone_start}. The values of the constants $\mu$ given in \eqref{eqn:sublayer_psdtraceone_start} and $\epsilon$ given in \eqref{eqn:epsilonintroduction} used for numerical stabilization of the gradients are $10^{-8}$ and $10^{-4}$ respectively. After training the network, we estimate the Bell state from the noisy POVM-probabilities obtained  by measuring the Bell state for $N_{avg}$ times. To see the performance of our network, fidelity and trace distance between the estimated density matrix and the Bell state are calculated. Results are averaged over 100 estimations. Following \cite{rqst_gm}, we also calculate the classic fidelity between the POVM-probabilities of the estimated density matrix ($P_{Est}(\bm{a})$) and the exact probabilities $P_{Bell}(\bm{a})$ which is defined below

\begin{definition}
    Classic fidelity between two probability mass functions (PMF) $P_{Bell}(\bm a)$ and $P_{Est}(\bm a)$ is defined as
    \begin{align}
             F_C( P_{Est}(\bm{a}) , P_{Bell}(\bm{a}) ) &= \mathbb{E}_{\bm{a} \sim  P_{Bell}(\bm{a}) } \bigg[ \sqrt{ \frac{ P_{Est}(\bm{a}) }{  P_{Bell}(\bm{a}) } } \bigg] \\
             &= \mathbb{E}_{\bm{a} \sim  P_{Est}(\bm{a}) } \bigg[ \sqrt{ \frac{ P_{Bell}(\bm{a}) }{  P_{Est}(\bm{a}) } } \bigg]
         \label{eqn:classicfidelity} \end{align} \label{def:cf}
\end{definition}

We estimate the Bell state by observing $m$ POVM-probabilities for different values of $m$. 1000 individual measurements are performed to calculate each POVM-probability ($N_{avg}=1000$). The corresponding results are plotted in Figure \ref{fig:bell_trace_diffm_T3_navg1000_noclr} . It is observed that with 10 POVM-probabilities, the network estimates Bell state with a Fidelity of 0.9176 and a classic fidelity of 0.9894 after 4000 epochs of training. When an informationally complete set of measurements is observed ($m=16$) our network estimates the Bell state with a fidelity of 0.9774 and classic fidelity of 0.9973.

To see how the value of $N_{avg}$ affects the performance, Bell state is estimated with different values of $N_{avg}$ and the corresponding results are plotted in Figure \ref{fig:bell_trace_diffnavg_m10_T3_noclr}. Since the size of the dataset used for training and validating is $M$, the number of individual measurements required for training can be calculated as $M \times N_{avg}$. The Bell state is estimated with informationally complete set of measurements and the corresponding results are plotted in Figures \ref{fig:bell_trace_diffnavg_m16_T3_noclr_fd}, \ref{fig:bell_trace_diffnavg_m16_T3_noclr_td} and  \ref{fig:bell_trace_diffnavg_m16_T3_noclr_cf}. It is observed that the network estimates Bell state with a very high fidelity of 0.9928 when $N_{avg}=5000$ is used. A high fidelity of 0.9284 and a classic fidelity of 0.9907 is achieved with only $N_{avg}=200$.  To see if the performance of our network increases with increasing $N_{avg}$ when informationally incomplete set of measurements are observed, we perform estimation with $m=10$ and the corresponding results are plotted in \ref{fig:bell_trace_diffnavg_m10_T3_noclr_fd}, \ref{fig:bell_trace_diffnavg_m10_T3_noclr_td} and  \ref{fig:bell_trace_diffnavg_m10_T3_noclr_cf}. Previously, when we varied the value of $m$, we observed that a fidelity of 0.9176 is achieved for $m=10$ when 1000 measurements are used to calculate POVM probabilities. This fidelity is increased to 0.9657 when $N_{avg}$ is increased to a high number of 10000. We observe that, even with an informationally incomplete set of measurements our network's performance is very good provided sufficient number of individual measurements are performed.

We would like to make a few comparisons between our network (LQST) and the RBM (Restricted Boltzmann Machine) network used by the authors of \cite{rqst_gm} which is again analyzed in \cite{nnqst2qubit} with the real-time dataset. In \cite{rqst_gm}, authors have used a generative model (RBM) to learn the POVM probabilities of the Bell state in an unsupervised manner. RBM learns the Bell state with a classic fidelity of 0.99 with 60,000 individual measurement outcomes. Note that 60,000 individual measurements correspond to $N_{avg}=100$ in our experiments since we use a dataset of 600 POVM-probabilities. With $N_{avg}=100$, LQST estimates the Bell state with a classic fidelity of 0.9831 from informationally complete measurements. The first difference between LQST and RBM is that RBM learns only a particular state (such as the Bell state) during the training process in an unsupervised manner whereas LQST is capable of learning a more general state in a supervised manner. For instance, we trained LQST to learn rank-1 states, and hence it can estimate the Bell state which is a rank-1 state. Once the training is done, being a generative model, RBM is capable of generating individual outcomes distributed as POVM-probabilities of that particular state which the network learned. LQST, on the other hand, is capable of reconstructing the density matrix when POVM probabilities are given as input to the network. RBM is trained by the outcomes of individual POVM measurements of a particular state (Bell state) while LQST is trained by the POVM probabilities. Since POVM-probabilities are computed from individual measurements, LQST demands more number of individual measurements be performed. A general LQST and the SVT algorithm require memory size exponential in the number of qubits. Unless a restriction on the number of parameters used in LQST is brought, LQST demands more memory size. Whereas, RBM requires less memory size than LQST and SVT. The number of parameters used in the RBM is the product of number of visible units and the number of hidden units. Since the number of visible units used is taken as the number of qubits, the memory size is atleast linear in the number of qubits. A trained RBM can only generate the POVM outcomes according to the learnt POVM probabilities. If the density matrix is to be reconstructed by inverting the POVM probabilities, then RBM must have learnt an informationally complete set of POVM probabilities. Further, to compute POVM-probabilities a finite number of POVM-outcomes must be drawn from the learnt RBM network. This leads to finite measurement error in POVM probabilities which in turn can lead to a non-valid density matrix (which violates the positive semidefiniteness condition) upon inversion whereas LQST can directly reconstruct the density matrix of a state from an informationally incomplete set of measurements of the state as we demonstrated through our numerical simulations.

\section{Conclusion} \label{sec:conclusion}

We designed and trained a deep neural network namely, LQST to perform quantum state tomography by unrolling a very popular compressive sensing algorithm called Singular Value Thresholding (SVT). We numerically simulated our algorithm to estimate the state of randomly generated 4-qubit systems and showed that our algorithm recovers the quantum state with better fidelity than SVT. To estimate a state with given fidelity LQST requires only a very few layers whereas SVT takes many thousands of iterations to converge. We demonstrated the estimation of 2 qubit quantum Bell state using LQST from an informationally incomplete set of noisy measurements. We compared our network LQST with a neural network-based quantum estimation method that uses the Restricted Boltzmann Machine (RBM). Though the memory used in RBM is much less compared with LQST and SVT, RBM learns only a particular quantum state in an unsupervised manner. LQST directly reconstructs the density matrix of a more general quantum state from an informationally incomplete set of measurements.

	\bibliographystyle{IEEEtran}
	\bibliography{reference.bib}

\begin{thebibliography}{10}
\providecommand{\url}[1]{#1}
\csname url@samestyle\endcsname
\providecommand{\newblock}{\relax}
\providecommand{\bibinfo}[2]{#2}
\providecommand{\BIBentrySTDinterwordspacing}{\spaceskip=0pt\relax}
\providecommand{\BIBentryALTinterwordstretchfactor}{4}
\providecommand{\BIBentryALTinterwordspacing}{\spaceskip=\fontdimen2\font plus
\BIBentryALTinterwordstretchfactor\fontdimen3\font minus
  \fontdimen4\font\relax}
\providecommand{\BIBforeignlanguage}[2]{{%
\expandafter\ifx\csname l@#1\endcsname\relax
\typeout{** WARNING: IEEEtran.bst: No hyphenation pattern has been}%
\typeout{** loaded for the language `#1'. Using the pattern for}%
\typeout{** the default language instead.}%
\else
\language=\csname l@#1\endcsname
\fi
#2}}
\providecommand{\BIBdecl}{\relax}
\BIBdecl

\bibitem{qcqi}
M.~A. Nielsen and I.~L. Chuang, \emph{Quantum Computation and Quantum
  Information: 10th Anniversary Edition}.\hskip 1em plus 0.5em minus
  0.4em\relax Cambridge University Press, 2010.

\bibitem{onthemeasofqubits}
D.~F.~V. James, P.~G. Kwiat, W.~J. Munro, and A.~G. White, ``Measurement of
  qubits,'' \emph{Physical Review A}, vol.~64, no.~5, Oct 2001.

\bibitem{mlmethodsinqst}
Z.~Hradil, J.~{\v{R}}eh{\'a}{\v{c}}ek, J.~Fiur{\'a}{\v{s}}ek, and
  M.~Je{\v{z}}ek, \emph{3 Maximum-Likelihood Methods in Quantum
  Mechanics}.\hskip 1em plus 0.5em minus 0.4em\relax Berlin, Heidelberg:
  Springer Berlin Heidelberg, 2004, pp. 59--112.

\bibitem{mlqst}
J.~B. Altepeter, D.~F. James, and P.~G. Kwiat, \emph{4 Qubit Quantum State
  Tomography}.\hskip 1em plus 0.5em minus 0.4em\relax Berlin, Heidelberg:
  Springer Berlin Heidelberg, 2004, pp. 113--145.

\bibitem{mlestofdensitymatrix}
K.~Banaszek, G.~M. D'Ariano, M.~G.~A. Paris, and M.~F. Sacchi,
  ``Maximum-likelihood estimation of the density matrix,'' \emph{Phys. Rev. A},
  vol.~61, p. 010304, Dec 1999.

\bibitem{dilutedml}
J.~Řeháček, Z.~Hradil, E.~Knill, and A.~I. Lvovsky, ``Diluted
  maximum-likelihood algorithm for quantum tomography,'' \emph{Physical Review
  A}, vol.~75, no.~4, Apr 2007.

\bibitem{bayesianinference}
D.~S. Gonçalves, C.~L.~N. Azevedo, C.~Lavor, and M.~A. Gomes-Ruggiero,
  ``Bayesian inference for quantum state tomography,'' \emph{Journal of Applied
  Statistics}, vol.~45, no.~10, pp. 1846--1871, 2018.

\bibitem{lowrankanybasisgross}
D.~{Gross}, ``Recovering low-rank matrices from few coefficients in any
  basis,'' \emph{IEEE Transactions on Information Theory}, vol.~57, no.~3, pp.
  1548--1566, 2011.

\bibitem{psdmf}
D.~Lahat, Y.~Lang, V.~Y.~F. Tan, and C.~Févotte, ``Positive semidefinite
  matrix factorization: A connection with phase retrieval and affine rank
  minimization,'' \emph{IEEE Transactions on Signal Processing}, vol.~69, pp.
  3059--3074, 2021.

\bibitem{lrmscaledsubgradient}
T.~Tong, C.~Ma, and Y.~Chi, ``Low-rank matrix recovery with scaled subgradient
  methods: Fast and robust convergence without the condition number,''
  \emph{IEEE Transactions on Signal Processing}, vol.~69, pp. 2396--2409, 2021.

\bibitem{powerfactorization}
J.~P. {Haldar} and D.~{Hernando}, ``Rank-constrained solutions to linear matrix
  equations using powerfactorization,'' \emph{IEEE Signal Processing Letters},
  vol.~16, no.~7, pp. 584--587, 2009.

\bibitem{als}
D.~{Zachariah}, M.~{Sundin}, M.~{Jansson}, and S.~{Chatterjee}, ``Alternating
  least-squares for low-rank matrix reconstruction,'' \emph{IEEE Signal
  Processing Letters}, vol.~19, no.~4, pp. 231--234, 2012.

\bibitem{sparesepowerfactorization}
K.~{Lee}, Y.~{Wu}, and Y.~{Bresler}, ``Near-optimal compressed sensing of a
  class of sparse low-rank matrices via sparse power factorization,''
  \emph{IEEE Transactions on Information Theory}, vol.~64, no.~3, pp.
  1666--1698, 2018.

\bibitem{rankonemeas}
Y.~{Li}, Y.~{Sun}, and Y.~{Chi}, ``Low-rank positive semidefinite matrix
  recovery from corrupted rank-one measurements,'' \emph{IEEE Transactions on
  Signal Processing}, vol.~65, no.~2, pp. 397--408, 2017.

\bibitem{psdrlm}
Q.~Zheng and J.~Lafferty, ``A convergent gradient descent algorithm for rank
  minimization and semidefinite programming from random linear measurements,''
  in \emph{Advances in Neural Information Processing Systems}, C.~Cortes,
  N.~Lawrence, D.~Lee, M.~Sugiyama, and R.~Garnett, Eds., vol.~28.\hskip 1em
  plus 0.5em minus 0.4em\relax Curran Associates, Inc., 2015.

\bibitem{permutationallyinvariant}
C.~Schwemmer, G.~T\'oth, A.~Niggebaum, T.~Moroder, D.~Gross, O.~G\"uhne, and
  H.~Weinfurter, ``Experimental comparison of efficient tomography schemes for
  a six-qubit state,'' \emph{Phys. Rev. Lett.}, vol. 113, p. 040503, Jul 2014.

\bibitem{qstsingleobservable}
D.~{Oren}, M.~{Mutzafi}, Y.~C. {Eldar}, and M.~{Segev}, ``Quantum state
  tomography with a single observable,'' in \emph{2017 Conference on Lasers and
  Electro-Optics (CLEO)}, 2017, pp. 1--1.

\bibitem{gross}
D.~Gross, Y.-K. Liu, S.~T. Flammia, S.~Becker, and J.~Eisert, ``Quantum state
  tomography via compressed sensing,'' \emph{Physical review letters}, vol.
  105, no.~15, p. 150401, 2010.

\bibitem{gross2}
C.~Riofr{\'\i}o, D.~Gross, S.~T. Flammia, T.~Monz, D.~Nigg, R.~Blatt, and
  J.~Eisert, ``Experimental quantum compressed sensing for a seven-qubit
  system,'' \emph{Nature communications}, vol.~8, no.~1, pp. 1--8, 2017.

\bibitem{qtprotocols}
A.~Kalev, R.~L. Kosut, and I.~H. Deutsch, ``Quantum tomography protocols with
  positivity are compressed sensing protocols,'' \emph{npj Quantum
  Information}, vol.~1, no.~1, pp. 1--6, 2015.

\bibitem{svt}
J.-F. Cai, E.~J. Cand{\`e}s, and Z.~Shen, ``A singular value thresholding
  algorithm for matrix completion,'' \emph{SIAM Journal on optimization},
  vol.~20, no.~4, pp. 1956--1982, 2010.

\bibitem{rqst_gm}
J.~Carrasquilla, G.~Torlai, R.~G. Melko, and L.~Aolita, ``Reconstructing
  quantum states with generative models,'' \emph{Nature Machine Intelligence},
  vol.~1, no.~3, pp. 155--161, mar 2019.

\bibitem{nnqst2qubit}
M.~Neugebauer, L.~Fischer, A.~J\"ager, S.~Czischek, S.~Jochim,
  M.~Weidem\"uller, and M.~G\"arttner, ``Neural-network quantum state
  tomography in a two-qubit experiment,'' \emph{Phys. Rev. A}, vol. 102, p.
  042604, Oct 2020.

\bibitem{unrollingsurvey}
V.~Monga, Y.~Li, and Y.~C. Eldar, ``Algorithm unrolling: Interpretable,
  efficient deep learning for signal and image processing,'' \emph{IEEE Signal
  Processing Magazine}, vol.~38, no.~2, pp. 18--44, 2021.

\bibitem{arm}
B.~Recht, M.~Fazel, and P.~A. Parrilo, ``Guaranteed minimum-rank solutions of
  linear matrix equations via nuclear norm minimization,'' \emph{SIAM review},
  vol.~52, no.~3, pp. 471--501, 2010.

\bibitem{boyd}
S.~Boyd, S.~P. Boyd, and L.~Vandenberghe, \emph{Convex optimization}.\hskip 1em
  plus 0.5em minus 0.4em\relax Cambridge university press, 2004.

\bibitem{uzawa}
K.~Arrow, H.~Chenery, H.~Azawa, K.~M.~R. Collection, L.~Hurwicz, H.~Uzawa,
  S.~Johnson, S.~Karlin, T.~Marschak, and R.~Solow, \emph{Studies in Linear and
  Non-linear Programming}, ser. Stanford mathematical studies in the social
  sciences.\hskip 1em plus 0.5em minus 0.4em\relax Stanford University Press,
  1958.

\bibitem{RPCA}
O.~{Solomon}, R.~{Cohen}, Y.~{Zhang}, Y.~{Yang}, Q.~{He}, J.~{Luo}, R.~J.~G.
  {van Sloun}, and Y.~C. {Eldar}, ``Deep unfolded robust pca with application
  to clutter suppression in ultrasound,'' \emph{IEEE Transactions on Medical
  Imaging}, vol.~39, no.~4, pp. 1051--1063, 2020.

\bibitem{adam}
D.~P. Kingma and J.~Ba, ``Adam: {A} method for stochastic optimization,'' in
  \emph{3rd International Conference on Learning Representations, {ICLR} 2015,
  San Diego, CA, USA, May 7-9, 2015, Conference Track Proceedings}, Y.~Bengio
  and Y.~LeCun, Eds., 2015.

\end{thebibliography}
\end{document}